\documentclass[11pt,a4paper]{article}
\pdfoutput=1
\usepackage{jheppub}
\usepackage{epsf,epsfig}
\usepackage{amscd}
\usepackage{amsmath}
\usepackage{epigraph}
\usepackage{hyperref}

\newcommand{\beq}{\begin{equation}}
\newcommand{\eeq}{\end{equation}}

\begin{document}

\title{Is Natural SUSY Natural?} 
\author[a]{Edward Hardy}
\emailAdd{e.hardy12@physics.ox.ac.uk}

\affiliation[a]{Rudolf Peierls Centre for Theoretical Physics,
University of Oxford,\\
1 Keble Road, Oxford,
OX1 3NP, UK}

\abstract{We study the fine tuning associated to a `Natural Supersymmetry' spectrum with stops, after RG running, significantly lighter than the first two generation sfermions and the gluino. In particular, we emphasise that this tuning should be measured with respect to the parameters taken to be independent at the assumed UV boundary of the renormalisation group flow, and improve the accuracy of previous approximate expressions. It is found that,  if running begins at $10^{16}~\rm{GeV}$ $\left(10^5~\rm{GeV}\right)$, decreasing the UV stop mass below $0.75$ $\left(0.4\right)$ of the weak scale Majorana gluino mass does not improve the overall fine tuning of the theory.  In contrast, it is possible to raise the first two generation sfermion masses out of LHC reach without introducing additional tuning. After running, regions of parameter space favoured by naturalness and consistent with LHC bounds typically have IR stop masses of order 1.5 TeV (0.75 TeV), and fine tuning of at least 400 (50) for high (low) scale mediation. We also study the fine tuning of theories with Dirac gluinos. These allow for substantial separation of the gluino and sfermion masses and, regardless of the scale of mediation, lead to relatively low fine tuning of order 50. Hence viable models can still favour light stops, but this requires extra structure beyond the MSSM field content.

}

\maketitle


\section{Introduction}
With the LHC giving increasingly strong limits on supersymmetric spectra with universal sfermion masses, models of `natural' supersymmetry (SUSY), where only superpartners directly involved in the tuning of the electroweak scale are light, provide an intriguing alternative \cite{Dimopoulos:1995mi,Cohen:1996vb}. Since the parton content of the proton means many production channels of supersymmetric particles are strongest through the first two generation sfermions, such spectra can relax collider limits dramatically and provide hope for an electroweak sector without significant fine tuning \cite{Kribs:2013lua,Krizka:2012ah,Auzzi:2012dv,Espinosa:2012in,Han:2012fw,Lee:2012sy,Bai:2012gs,Allanach:2012vj,Larsen:2012rq,Bi:2011ha,Brust:2011tb,Arganda:2012qp,Arganda:2013ve,Cao:2012rz}. However, as was quickly realised after their initial proposal, it is difficult to preserve a natural spectrum, which requires light stops, during running to the weak scale \cite{ArkaniHamed:1997ab,Hisano:2000wy}. On one hand, the heavy first two generation sfermions tend to drive the stops tachyonic, while on the other, a gluino above the current experimental limit will tend to pull the stops to unacceptably high masses. These running effects manifest themselves in the electroweak sector as  two loop contributions to the up type Higgs' soft mass.

Quantifying the fine tuning of a model is a useful tool to study the viability of particular low energy spectra \cite{Barbieri198863}. This approach has been applied in a large number of studies of supersymmetric models, for example \cite{Kitano:2006gv,Kitano:2005wc,Antusch:2012gv,Feng:2013pwa,Arvanitaki:2012ps,Strumia:1999fr,Kane:1998im,Kang:2012sy,Cassel:2010px,Cassel:2009cx,Ghilencea:2013fka,Athron:2013ipa,Cabrera:2008tj,Casas:2003jx,deCarlos:1993yy,Badziak:2012rf,Ross:2011xv},  has been used to strongly constrain spectra with universal sfermion masses, and has also been studied in the context of natural spectra \cite{Baer:2012up,Papucci:2011wy,Agashe:1998zz,Wymant:2012zp,Ghilencea:2012qk}. In this paper, we first derive expressions for the fine tuning required to obtain stops significantly lighter than gluinos and the first two generation sfermions. We then extend previous approximate results for the fine tuning of the electroweak scale introduced due to heavy gluinos and sfermions.  Applying current experimental contraints these are used to study the extent to which fine tuning may be evaded. Our main result is that if there is a Majorana gluino with mass $>1.5 \, \rm{TeV}$ there is no fine tuning benefit to decreasing the stop masses below roughly 1 TeV if mediation is from close to the GUT scale. This is because such theories necessarily contain a significant amount of fine tuning in the electroweak sector from the gluino feeding into the Higgs mass at two loops. However, while there is no benefit to reducing the stop mass, provided the stop is not too light ($\gtrsim 500 \, \rm{GeV}$) doing so does not actually make the tuning of the theory worse and is not actively disfavoured. As a result of this, we are able to put a strong lower bound on the fine tuning of theories of natural SUSY, even though there are  regions of parameter space where the LHC has not excluded light stops.

An important point for our study is that we assume particular renomalisation group boundary conditions at some energy scale. The fine tuning of the theory is then measured with respect to the parameters of the theory at this boundary, which are assumed to be independent.\footnote{Note however, the choice of the location of this boundary, and the set of independent parameters there, is only physically meaningful once a complete UV theory, including all higher dimension operators, is specified.} In contrast, the weak scale parameters have values which are strongly coupled together by the renormalisation group equations and attempting to quantify the fine tuning of a theory in terms of them has the capacity to miss important effects from running (the importance of this has been emphasised in recent papers \cite{Baer:2012mv,Baer:2013bba}). Of course, choosing the independent variables at the UV boundary automatically requires some assumptions about the mediation of supersymmetry breaking, and in particular possible correlations between soft terms at this scale. Additionally, we must assume there is no new physics between the UV boundary and the weak scale that modifies the running, the possibility that this assumption does not hold due to interactions in the SUSY breaking sector has been studied in \cite{Dine:2004dv,Cohen:2006qc}. For the majority of our study, we take the independent variables to be the gluino mass (the other gauginos are less important for our study and we do not need to assume a GUT structure), the stop mass and the mass of the first two generation sfermions, which are assumed to be universal based on strong flavour constraints \cite{Contino:1998nw}.\footnote{Though see, for example, \cite{Kribs:2007ac,Abdullah:2012tq,Perez:2012mj,Mahbubani:2012qq,Galon:2013jb} for a discussion of ways in which this assumption may be relaxed.} Such a choice is reasonable; to obtain a natural spectrum typically requires boundary conditions with  heavy first two generation sfermions, an intermediate mass gluino and stops with masses somewhat, but not too far, below the gluino. This is usually accomplished by including several mediation mechanisms which couple to different visible sector states. For example, the first two generation sfermions may gain their mass dominantly through a D-term of an additional U(1) gauge group \cite{Dvali:1996rj,Nelson:1997bt,Kaplan:1998jk,Kaplan:1999iq,Craig:2012di,Hardy:2013uxa}, while the gluino and stops gain their mass either through another form of gauge mediation or gravity mediation. Hence, these masses may be adjusted independently. Additionally, in both gravity mediation \cite{Brignole:1997dp} and the most general models of gauge mediation \cite{Meade:2008wd}, the gauge fermion and sfermion masses generated are independent.

There is an alternative scenario which is also well motivated. Suppose, the gluino and stop masses at the UV renormalisation boundary are both generated through a single F-term, as the result of an especially simple SUSY breaking sector and mediation mechanism. Now, varying the gluino mass will be correlated to varying the UV stop mass, and hence we should take the F-term to be our fundamental parameter. As we will discuss later, this scenario actually makes the tuning of natural SUSY spectra substantially worse since increasing the F-term increases the weak scale stop mass both directly though the UV stop mass, and through the increased running from a more massive gluino. In this way our study can be seen as providing a lower bound on the fine tuning obtained. A more serious question is whether the left and right handed stop masses should be regarded as one parameter, as is the case if both gain the majority of their soft mass through the same mediation mechanism. This is expected to be the case in many models of natural SUSY, however is not required in generic mediation. We give results for the both the case where these are independent, and when they are not.

There is a possible proviso to our argument. It might be the case that the mediation mechanism somehow favours UV spectra which, as the magnitude of the SUSY breaking is varied, preserves a particular structure which minimises the running (this is the case for focus point spectra \cite{Feng:1999zg,Horton:2009ed}). However, such a mechanism would need to couple the stop, gluino, and first two generation sfermions in a highly non-trivial way despite their soft masses coming from very different sources (typically R-symmetry preserving SUSY breaking, R-symmetry breaking SUSY breaking and an additional D-term respectively), and there seems to be no reason that SUSY breaking and mediation should know anything at all about the MSSM renormalisation group equations. Therefore, this does not seem a strong assumption.\footnote{In contrast focus point scenarios typically only involve one, simple, form of mediation to all MSSM fields, hence can occur as a result of single numerical coincidence in the structure of the mediators which seems far less artificial than would be required for a natural SUSY spectrum.}

As a final caveat of our work, we have studied only the \emph{sensitivity} of the electroweak scale to the UV parameters. We make no attempt to quantify the probability, over the theory space of SUSY breaking and mediation mechanisms, that the initial UV parameters begin in the correct region to allow for a natural spectrum at the weak scale. Since, as discussed, such a starting point requires multiple forms of mediation which, \emph{a priori}, could lead to a separation between the gluino and sfermion masses which is far too large to lead to a viable natural spectrum at the weak scale. Hence, it may be thought that natural spectra are rare over the space of models. However, there may be some hope in this direction by linking the ratio of gluino to first two generation sfermion masses to another parameter of approximately the correct size in the model, for example the parameter $\xi^2$ in string theory or the ratio of fermion masses \cite{Hardy:2013uxa,Craig:2012di}.

While the main focus of our work is on conventional Majorana gauginos, an interesting alternative is to introduce additional fields that allow the generation of Dirac gaugino mass term. We study the electroweak fine tuning in a simple example of such a model, and find that, independent of the mediation scale, it is comparable to a MSSM theory with very low cutoff. Hence, this is a good option for reducing fine tuning in models where the mediation scale is required to be high, for example if attempting to build a string-motivated UV completion.

As is well known, there is also a tension between light stops and the observed Higgs mass of $\sim 125\,\rm{GeV}$. At tree level in the MSSM, the Higgs mass is bounded by the mass of the Z boson, and radiative corrections from fairly heavy stops are required to raise its mass to the observed value (see, for example, \cite{Martin:1997ns}). For the purposes of this work, we assume this can be evaded through an NMSSM like model, in which an additional singlet is present giving an extra tree level contribution to the Higgs mass. The extra field content of such a model does not alter the leading dependence of the Higgs mass on the gluino, stops, and sfermions during running so will not affect our fine tuning results, and is independently motivated for its ability to solve the $\mu$ problem  \cite{Ellwanger:2009dp}. The extra field content will somewhat change the fine tuning with respect to the soft Higgs mass, which we calculate within the MSSM, however the parametric form will be unchanged, and ultimately we will find this is not typically the dominant tuning. Even in the NMSSM very light stops are potentially problematic,  since in this case, the coupling, $\lambda$, of the singlet, $S$, to the Higgs through the term $\lambda S H_u H_d$ must be large at the weak scale \cite{Barbieri:2006bg}. Typically such values, lead to $\lambda$ running to a strong coupling regime before $10^{16}\,\rm{GeV}$, although this does not necessarily ruin the successful prediction of gauge unification \cite{Hardy:2012ef}. In contrast, we will find that the most natural regions of parameter space not yet excluded by LHC limits may have relatively heavy stop masses, which allow $\lambda$ to be small or the Higgs mass to be generated directly in the MSSM without additional structure.

Turning to the structure of this paper, in Section \ref{sec:tuning} we discus the fine tuning of the UV parameters required to obtain a light stop after running. Section \ref{sec:ew} contains the main results on the tuning of the electroweak VEV in natural scenarios, while Section \ref{sec:dirac} contains our discussion of Dirac gauginos.


\section{Fine Tuning to Obtain a Light Stop} \label{sec:tuning}
We begin by briefly reviewing the fine tuning of the electroweak scale introduced by stops, as was defined in the early phenomenological studies of supersymmetry \cite{Barbieri198863}. The fine tuning due to the weak scale values of the stops is given by
\begin{equation} \label{eq:ft1}
\tilde{Z}_{\tilde{t}} =\biggl| \frac{\partial\left(\log M_Z^2\right)}{\partial\left(\log m_{\tilde{t}}^2\left(M_W\right)\right)}\biggr| = \frac{m_{\tilde{t}}^2}{M_Z^2} \frac{\partial M_Z^2}{\partial m_{\tilde{t}}^2\left(M_W\right)} ,
\end{equation}
where, for future convenience, the tilde denotes that this is a fine tuning with respect to the theory's weak scale parameters. We will generally use the convention that soft terms without their scale specified are evaluated at the UV boundary of the renormalisation flow of the theory, $\Lambda_{UV}$, which is typically the scale at which SUSY breaking is mediated.

It is straightforward to estimate $\tilde{Z}_{\tilde{t}} $. Stops give a contribution to the up type Higgs through running, which is given at leading log level by
\begin{equation}
\delta m_{H_u}^2\left(M_W\right) = \frac{-3 y_t^2}{8 \pi^2} \left(m_{u3}^2\left(M_W\right)+ m_{Q3}^2\left(M_W\right)+A_t^2\left(M_W\right)\right)\log\left(\frac{\Lambda_{UV}}{m_{\tilde{t}}}\right) .
\end{equation}
Additionally, to a good approximation,
\begin{equation}
\tilde{Z}_{m_{\tilde{Q3}}} = \biggl|\frac{2 \delta m_{H_u}^2}{M_Z^2}\biggr| . \label{eq:zuphiggs}
\end{equation}
Hence the fine tuning parameter at this order is
\begin{equation}
\tilde{Z}_{m_{\tilde{Q3}}} =  \frac{3}{4 \pi^2 \cos\left(2 \beta \right)} \frac{m_t^2}{v^2  M_Z^2} \log\left(\frac{\Lambda_{UV}}{m_{\tilde{t}}}\right)  m_{\tilde{Q3}}^2\left(M_W\right) . \label{eq:sft}
\end{equation}
Normally, the parameters $\tilde{Z}_i$ for all of the variables $i$ are compared, and the overall fine tuning is defined as $\rm{max}\left(\{\tilde{Z}_i\}\right)$. However, we will focus on the fine tuning introduced through stops, gluinos and the first two generation sfermions since these are the tunings which are relevant for considering natural spectra. In Section \ref{sec:ew} we compute the electroweak fine tuning with respect to the UV values of these parameters to a higher order. In particular, this is necessary because we will be interested in running from the GUT scale, and in this case the expansion parameter is $\frac{b_3 \alpha_3}{2\pi^2}\log\left(\frac{10^{16}}{10^3}\right) \sim 0.5$, which is not especially small. In complete theories, there are other important tunings due to the $\mu$ and $B\mu$ parameters, and depending on the details of the electroweak sector soft terms these may be significant. Therefore, our analysis will give a \emph{lower bound} on the fine tuning.

Now we turn to the fine tuning of the gluino and first two generation masses required to obtain a light stop at the weak scale. This is defined as
\begin{align}
Y_{i} =  \biggl|\frac{\partial\left(\log m_{\tilde{t}}^2\left(M_W\right)\right)}{\partial\left(\log i \right)}\biggr|  ,
\end{align} 
where $i$ is one of $M^2_3$, $\tilde{m}^2_{1,2}$ or $m_{\tilde{t}}^2$ evaluated at the UV boundary, and $\tilde{t}$ is the stop state which receives the greatest fine tuning.

The renormalisation group equations for the stops in the presence of heavy sfermions are well known, for example from \cite{ArkaniHamed:1997ab,Martin:1997ns}. Since we are interested in the effect of the gluino and sfermion masses and these dominate the renormalisation group equation, it is sufficient to include only the leading effects. Later we will see the next corrections are small. The running is given by
\begin{equation}
\frac{d}{dt} m_{\tilde{t}}^2 = -\frac{8}{4\pi} \sum \alpha_i\left(t\right) C_i M_i^2 + \frac{2}{\pi^2} \left( \sum \alpha_i^2\left(t\right) C_i \right) \tilde{m}_{1,2}^2 ,
\end{equation}
where $C_i$ is the Casimir of the stop state (and $\alpha_1$ is GUT normalised). We further assume the right handed bottom sfermion and the staus remain relatively light such that they do not have a significant effect on the running of the stops, but not so light as to be driven tachyonic during running. Giving these states a large mass would in general make the running faster and the fine tuning worse. This is not a large effect and does not significantly alter any of our conclusions. We take the heavy first two generations to have a constant mass which is a reasonable approximation if they begin fairly heavy as in natural spectra (in our numerical analysis we include the subleading effect from their running).\footnote{We are interested in spectra where the stops are fairly light at the UV scale and remain relatively light during running. Hence, the overall shift in their mass during running is $\lesssim 500 \, \rm{GeV}$. The first two generation's dominant running is the same as the stops hence these run by a similar amount, which is negligible if they start at $\mathcal{O}\left(10 \, \rm{TeV}\right)$.} Following   \cite{ArkaniHamed:1997ab}, at this level of approximation the flow can be solved exactly to give
\begin{equation}
\begin{aligned}
m_{\tilde{t}}^2\left(M_W\right) = &m_{\tilde{t}}^2\left(\Lambda_{UV} \right)  -\sum_i \frac{2}{b_i} C_i \left(\frac{1}{\left(1 + \frac{b_i}{2\pi} \log\left(\frac{\Lambda_{UV}}{M_i\left(M_W\right)}\right)\alpha_i\right)^2}-1\right) M_i^2\\
&\qquad \qquad +\sum_i \frac{4}{\pi b_i} \alpha_i\left(\Lambda_{UV} \right) \left( \frac{1}{1+\frac{b_i}{2\pi}\log\left(\frac{\Lambda_{UV}}{\tilde{m}_{1,2}\left(M_W\right)}\right)\alpha_i}-1\right)C_i \tilde{m}_{1,2}^2 ,  \label{eq:stoprunning}
\end{aligned}
\end{equation}
where the gauge beta-function coefficients are defined as $\frac{d}{dt} \left(\frac{1}{\alpha_i}\right) = -\frac{b_i}{2\pi}$. Note that the contribution from the first two generation sfermion turns off at an energy scale $\tilde{m}^2_{1,2}$ while the gaugino contribution is present until the scale $M_i$. Equation \eqref{eq:stoprunning} is written in terms of the UV values of the gauge couplings to avoid an extra term when varying with respect to the soft masses. 

\begin{figure}[t!]
\begin{center}
\includegraphics[width=140mm]{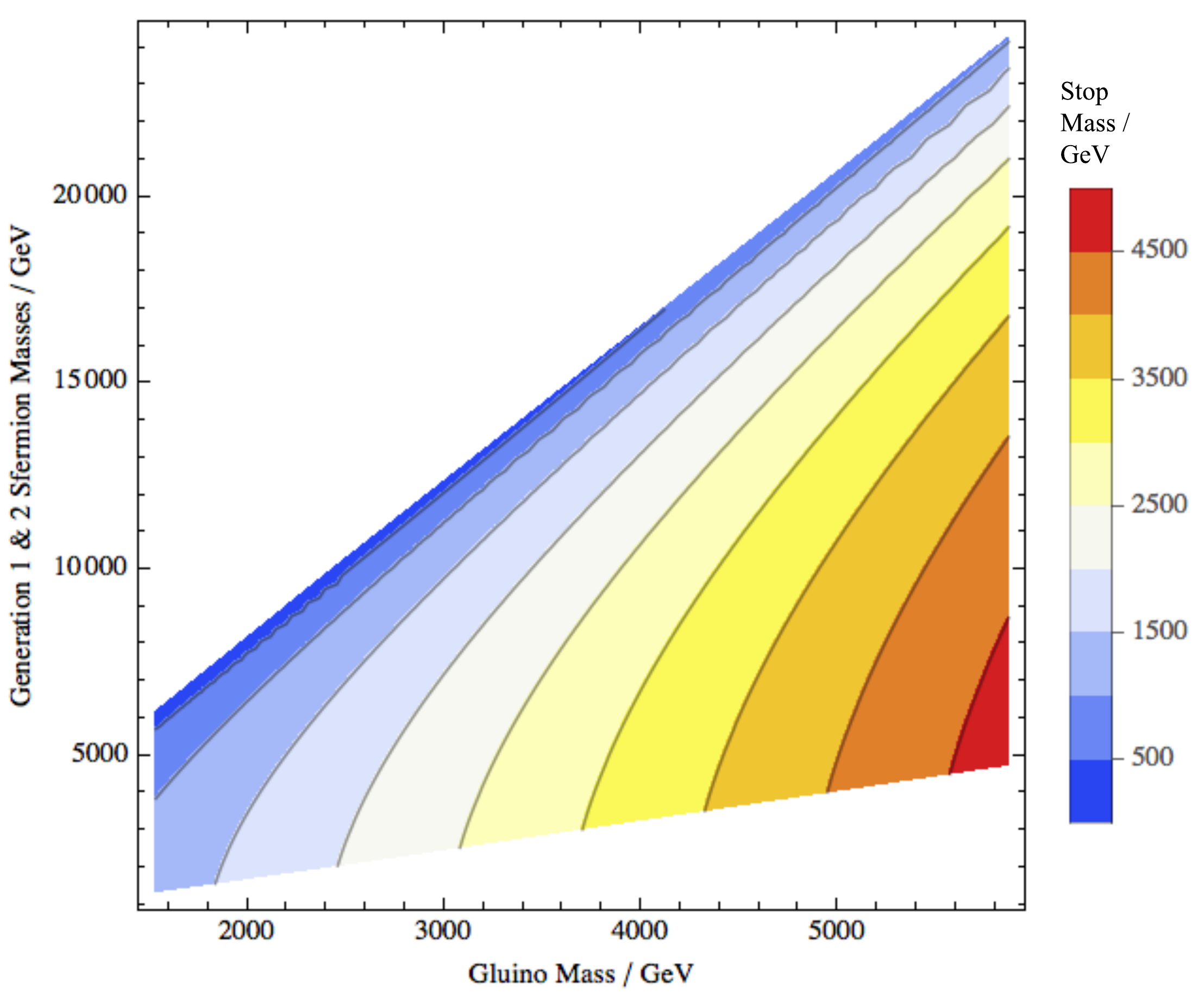}
\caption{The stop mass obtained at the weak scale as a function of the weak scale gluino and first two generation sfermion masses, after running from the GUT scale at $10^{16} \, \rm{GeV}$ assuming an initial mass of $200 \, \rm{GeV}$. The lower cutoff is due to the gluino increasing the the first two generation sfermions masses during running, while the upper cutoff is due to the stop running tachyonic above this line. }
\label{fig:1}
\end{center}
\end{figure}

In Fig.\ref{fig:1} we plot the weak scale lightest stop mass as a function of the weak scale gluino and first two generation sfermion masses, after running from $10^{16}\,\rm{GeV}$ with a UV stop mass of $200\,\rm{GeV}$. This shows that, for a given gluino mass, above a certain sfermion mass the stops run tachyonic and there is no viable electroweak spectrum. To obtain the light stops needed for a natural spectrum requires $M_3$ and $\tilde{m}_{1,2}$ to be such that the stop is in the thin strip close to this boundary.  The relatively small effect of the gluino increasing the mass of the first two generation sfermions during running is visible in the lower cut off in this plot.

Now it is straightforward to write down the fine tuning with respect to the UV gaugino and first two generation masses. There will be two contributions to the fine tuning, one directly from the dependence on $\tilde{m}_{1,2}^2$, and the other from dependence inside the logarithm,
\begin{equation}
\begin{aligned} \label{eq:ftm12}
Y_{\tilde{m}_{1,2}^2} &= \frac{\tilde{m}_{1,2}^2}{m_{\tilde{t}}^2}  \frac{\partial m_{\tilde{t}}^2}{\partial {\tilde{m}_{1,2}^2}} \\
&= \frac{\tilde{m}_{1,2}^2}{m_{\tilde{t}}^2}  \sum_i  \frac{4C_i}{\pi b_i} \alpha_i\left(\Lambda_{UV} \right) \left( \frac{1}{1+\frac{b_i}{2\pi} \log\left(\frac{\Lambda_{UV}}{\tilde{m}_{1,2}}\right)\alpha_i}-1\right)     \\
& \qquad \qquad \qquad \qquad +  \frac{\tilde{m}_{1,2}^2}{m_{\tilde{t}}^2} \sum_i \frac{C_i \alpha_i^2\left(\Lambda_{UV}\right)}{\pi^2}  \left( \frac{1}{1+\frac{b_i}{2\pi} \log\left(\frac{\Lambda_{UV}}{\tilde{m}_{1,2}}\right) \alpha_i}           \right)^2    .
\end{aligned}
\end{equation}
The second term from the variation of the logarithm is typically significantly smaller than the first and acts to reduce the fine tuning. This is expected, if the mass of the first two generation sfermions increases then there will be slightly less running.
 Similarly, we find
 \begin{equation}
\begin{aligned} \label{eq:YM3}
Y_{M_i^2\left(M_W\right)} &= - \frac{ M_i^2}{m_{\tilde{t}}^2} \frac{2}{b_i} C_i \left(\frac{1}{\left(1 + \frac{b_i}{2\pi} \log\left(\frac{\Lambda_{UV}}{M_i\left(M_W\right)}\right)\alpha_i\right)^2}-1\right) \\[1em]
&\qquad \qquad \qquad \qquad  -  \frac{ M_i^2}{m_{\tilde{t}}^2}   \frac{C_i}{\pi}  \alpha_i\frac{1}{\left(1 + \frac{b_i}{2\pi} \log\left(\frac{\Lambda_{UV}}{M_i\left(M_W\right)}\right)\alpha_i\right)^3} .
\end{aligned}
\end{equation}
It is clear that the greatest fine tuning from the heavy sfermions will occur on the left handed stop. This is because, even though the beta function coefficients $b_2$ and $b_3$ have opposite signs, their overall contributions to \eqref{eq:ftm12} go in the same direction. For the tuning with respect to the gauginos, we focus on the gluino, which couples equally to the left and right handed stops, since this is clearly dominant.

Finally, there is also a fine tuning with respect to the initial stop masses. This can be evaluated as a perturbation to the trajectory obtained already. As discussed in the Introduction, it is unclear if the left and right handed stops should be treated as independent variables. If the masses are independent, a perturbation to the initial left handed soft mass,  $\Delta m^2_{\tilde{Q3}}$, will satisfy
\begin{equation}
\frac{d}{dt}\left(\Delta m^2_{\tilde{Q3}} \right)\supset \frac{2 y_t^2}{16 \pi^2} \Delta m^2_{\tilde{Q3}} ,
\end{equation}
and will also feed into the right handed soft and up type Higgs mass since the renormalisation group includes
\begin{align}
\frac{d}{dt}\left(\Delta m^2_{\tilde{u3}} \right)&\supset \frac{4 y_t^2}{16 \pi^2} \Delta m^2_{\tilde{Q3}} ,\\
\frac{d}{dt}\left(\Delta m^2_{Hu} \right)&\supset \frac{6 y_t^2}{16 \pi^2} \Delta m^2_{\tilde{Q3}} .
\end{align}
Since the beta functions are linear in $m_{\tilde{t}}^2$, the evolution of the perturbation during running may be obtained by integrating the full one loop renormalisation equations (assuming MSSM field content and interactions)
\begin{align}
\Delta m^2_{\tilde{Q3}}\left(M_W\right) &=  \frac{1}{6} \left(5 + \left(\frac{m_{\tilde{Q3}}}{\Lambda_{UV}}\right)^{\left(3 y_t^2/4\pi^2 \right)}\right) \Delta m^2_{\tilde{Q3}}\left(\Lambda_{UV}\right) , \label{eq:a1}\\
\Delta m^2_{\tilde{u3}}\left(M_W\right) &=  \frac{1}{3} \left(-1+\left(\frac{m_{\tilde{Q3}}}{\Lambda_{UV}}\right)^{\left(3 y_t^2/4\pi^2 \right)}\right) \Delta m^2_{\tilde{Q3}}\left(\Lambda_{UV}\right) \label{eq:stopt} .
\end{align}
Similarly, a perturbation to the right handed stop gives
\begin{align}
\Delta m^2_{\tilde{u3}}\left(M_W\right) &=  \frac{1}{3} \left(2+\left(\frac{m_{\tilde{u3}}}{\Lambda_{UV}}\right)^{\left(3 y_t^2/4\pi^2 \right)}\right) \Delta m^2_{\tilde{u3}}\left(\Lambda_{UV}\right) ,  \label{eq:a2} \\
\Delta m^2_{\tilde{Q3}}\left(M_W\right) &= \frac{1}{6} \left(-1+\left(\frac{m_{\tilde{u3}}}{\Lambda_{UV}}\right)^{\left(3 y_t^2/4\pi^2 \right)}\right) \Delta m^2_{\tilde{u3}}\left(\Lambda_{UV}\right) .
\end{align}
Numerically, the expressions \eqref{eq:a1} and \eqref{eq:a2} dominate. therefore, the fine tunings are approximately
\begin{align}
Y_{m_{\tilde{Q3}}^2}&=\frac{m_{\tilde{Q3}}^2\left(\Lambda_{UV}\right)}{m_{\tilde{Q3}}^2\left(M_W\right)}  \left(\frac{5}{6}+\frac{1}{6}\left(\frac{m_{\tilde{Q3}}}{\Lambda_{UV}}\right)^{\left(3 y_t^2/4\pi^2 \right)}\right)  ,\\
Y_{m_{\tilde{u3}}^2}&=\frac{m_{\tilde{u3}}^2\left(\Lambda_{UV}\right)}{m_{\tilde{u3}}^2\left(M_W\right)}  \left(\frac{1}{3}\left(\frac{m_{\tilde{u3}}}{\Lambda_{UV}}\right)^{\left(3 y_t^2/4\pi^2 \right)}+\frac{2}{3}\right)  .
\end{align}
The behaviour of these expressions is interesting. If there is a small separation between the mediation scale and the weak scale then $  \left(\frac{m_{\tilde{Q3}}}{\Lambda_{UV}}\right)^{\left( 3y_t^2/4\pi^2 \right)} \sim 1 $ and the fine tuning $Y_{m_{\tilde{Q3}}^2} \sim \frac{m_{\tilde{Q3}}^2\left(\Lambda_{UV}\right)}{m_{\tilde{Q3}}^2\left(M_W\right)}  $ as is the leading order expectation. However if there is a large separation between these scales then running proceeds for sufficiently long that the stop back reaction from a perturbation suppresses the initial perturbation, reducing the fine tuning. For a mediation scale of $10^{16} \, \rm{GeV}$,
\begin{equation}
\left(\frac{m_{\tilde{Q3}}}{\Lambda_{UV}}\right)^{\left( 3y_t^2/ 4\pi^2 \right)} \sim 0.1,
\end{equation}
so this can be a significant effect in the models we are interested in. Not surprisingly the tuning of the left handed stop is greater since it is less strongly damped by the renormalisation.

In the case where these two stop masses are linked, the renormalisation group equations for the perturbation are modified since the left handed stop perturbation feeds into the right handed stop perturbation and vice versa. These are easily integrated to obtain
\begin{align}
\Delta m^2_{\tilde{Q3}}\left(M_W\right) &= \frac{1}{3} \left(\left(\frac{m_{\tilde{t3}}}{\Lambda_{UV}}\right)^{\left(3 y_t^2/4\pi^2 \right)}+2\right) \Delta m^2_{\tilde{t3}}\left(\Lambda_{UV}\right) , \\
\Delta m^2_{\tilde{u3}}\left(M_W\right) &= \frac{1}{3} \left(2\left(\frac{m_{\tilde{t3}}}{\Lambda_{UV}}\right)^{\left( 3 y_t^2/4\pi^2 \right)}+1\right) \Delta m^2_{\tilde{t3}}\left(\Lambda_{UV}\right) .
\end{align}
Therefore,
\begin{equation}
Y_{m_{\tilde{t3}}^2} = \frac{m_{\tilde{t3}}^2\left(\Lambda_{UV}\right)}{m_{\tilde{Q3}}^2\left(M_W\right)} \left(\left(\frac{m_{\tilde{t3}}}{\Lambda_{UV}}\right)^{\left( 3 y_t^2/4\pi^2 \right)}+1\right).
\end{equation}
As before, for $\Lambda_{UV}$ not too large, the damping is not significant and these expressions reduce to the leading order expectation $Y_{m_{\tilde{t3}}^2} \sim \frac{m_{\tilde{t3}}^2\left(\Lambda_{UV}\right)}{m_{\tilde{Q3}}^2\left(M_W\right)}$. However if $\Lambda_{UV}$ is close to the GUT scale the difference can be significant.

It is useful to gain some physical insight by finding approximate expressions in various limits. For $\Lambda_{UV}=10^{16}\, \rm{GeV}$, \eqref{eq:ftm12} and \eqref{eq:YM3} reduce to
\begin{align}
Y_{\tilde{m}_{1,2}^2} &\simeq  0.03 \frac{\tilde{m}_{1,2}^2}{m_{\tilde{t}}^2} \\
Y_{M_3^2} &\simeq 0.74 \frac{ M_3^2\left(M_W\right)}{m_{\tilde{t}}^2} ,
\end{align}
where the sfermion and stop masses are evaluated at the UV boundary. For a low scale model with $\Lambda_{UV}=10^{6}\, \rm{GeV}$ we obtain,
\begin{align}
Y_{\tilde{m}_{1,2}^2} &\simeq  0.0079 \frac{\tilde{m}_{1,2}^2}{m_{\tilde{t}}^2} \\
Y_{M_3^2} &\simeq 0.36 \frac{ M_3^2\left(M_W\right)}{m_{\tilde{t}}^2}  .
\end{align}

Of course, the fine tuning is significantly smaller in the low scale case. In the high scale case the stop will tend to be pulled up to within $\sqrt{0.74}\sim0.9$ of the gluino mass, while in the low scale case the stop will be pulled to $\sim 0.6$ of the gluino mass. The first two generation sfermions have a smaller effect, typically decreasing the stop masses by an amount given by $\sim 0.2$ and $\sim 0.08$ of their mass in the high and low scale mediation respectively. Also, the subleading correction from the variation of the logarithm is more important in models of low scale mediation, which matches intuition. These are found to agree within $\sim20\%$ with the variations evaluated numerically using the code \textsc{SOFTSUSY} \cite{Allanach:2001kg}. The next correction term is due to the back reaction from the contribution to $m_{\tilde{t}}$ to its own renormalisation group equation. In Appendix \ref{app:backreaction}, we compute this effect and include it in our numerical work.

In order to gauge the severity of these fine tunings, recall the expression for the tuning of the electroweak scale (we take $\cos\left(2\beta\right)=1$ which gives a minimum value for the tuning),
\begin{align}
\tilde{Z}_{\tilde{Q}3} \sim  \tilde{Z}_{\tilde{u}3} \sim  9.1\times 10^{-6}m_{\tilde{Q}3}^2/\rm{GeV}^2 \log\left(\frac{\Lambda_{UV}}{m_{\tilde{t}}} \right) .
\end{align}
A minimal SUGRA spectrum with sfermions and gluinos at $2500\, \rm{GeV}$ would have a $Z_{\tilde{Q}3}\sim 350$ for $\Lambda_{UV}=10^{6}\, \rm{GeV}$ and $Z_{\tilde{Q}3}\sim 1500$ for $\Lambda_{UV}=10^{16}\, \rm{GeV}$. In contrast, a reasonable natural spectra  has $m_{\tilde{t}}=200\, \rm{GeV}$, $\tilde{m}_{1,2}\sim 10^4 \,  \rm{GeV}$ and $M_3 \sim 2500 \, \rm{GeV}$ at the weak scale. If  $\Lambda_{UV}=10^{16}\,\rm{GeV}$, we obtain
\begin{align}
 Y_{\tilde{m}_{1,2}^2}\sim 80, \quad Y_{M_3^2} \sim 115,\quad Y_{m^2_{\tilde{t}3}}\sim 15, \quad Y_{m^2_{\tilde{Q}3}}\sim 20. 
 \end{align}
For a low scale model, with $\Lambda_{UV}=10^{6}\, \rm{GeV}$\begin{align}
Y_{\tilde{m}_{1,2}^2}\sim 20, \quad Y_{M_3^2} \sim 50, \quad  Y_{m^2_{\tilde{t}3}}\sim 20, \quad Y_{m^2_{\tilde{Q}3}}\sim 25. 
\end{align} 

\begin{figure}[t!]
\begin{center}
\includegraphics[width=160mm]{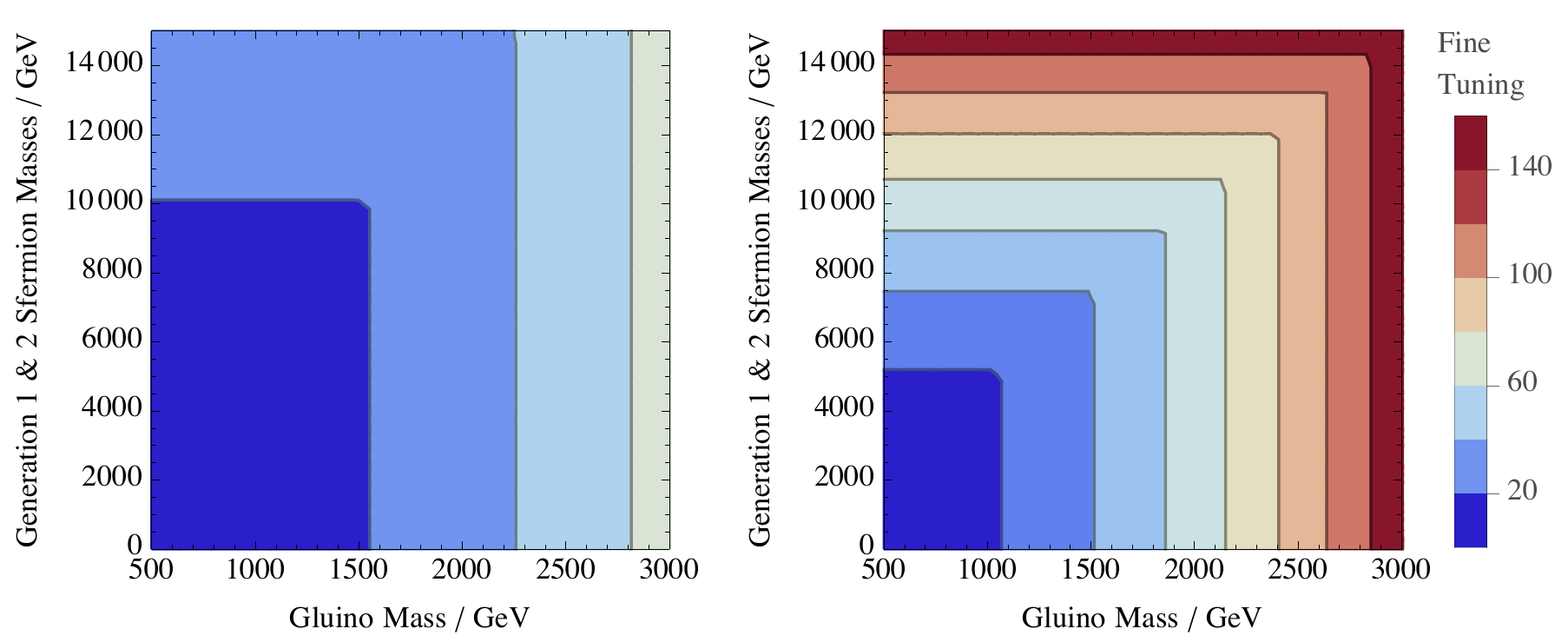}
\caption{The fine tuning required to obtain a stop mass of $200 \, \rm{GeV}$ at the weak scale, with {\bf Left:} A UV boundary of $10^{16} \, \rm{GeV}$ and {\bf Right:} A UV boundary of $10^{6} \, \rm{GeV}$, as a function of the \emph{weak scale} gluino mass and the UV value of the first two generation sfermion masses.}
\label{fig:sft}
\end{center}
\end{figure}

We define the overall fine tuning $Y$ to be given by $Y=\rm{max}\left(\{Y_{M_3^2}, Y_{\tilde{m}^2_{1,2}},Y_{m_{t3}^2}\}\right)$. In Fig.\ref{fig:sft} we show the fine tuning required to obtain a stop of mass $200 \, \rm{GeV}$ at the weak scale in the plane $M_3\left(M_W\right)$, $\tilde{m}_{1,2}$ evaluated at the weak scale for high and low scale mediation, assuming the two stop masses are not independent in the UV. These examples demonstrate our first result, it is possible to obtain a fairly light stop in the presence of a gluino mass $> 2 \, \rm{TeV}$ and first two generation sfermions with mass $> 5 \, \rm{TeV}$ with a tuning of order $5 \div 100$ depending on the scale of mediation. By itself this is not a large tuning compared to that found in the electroweak sector of typical MSSM models or extensions, hence a light stop should not be regarded as a particularly tuned scenario in itself. 


\section{Electroweak Fine Tuning in Models of Natural SUSY} \label{sec:ew}
We now turn to the question of whether a natural SUSY scenario, compatible with current limits, can lead to an electroweak sector with low fine tuning. For simplicity, we assume the MSSM parameters and interactions and that $\tan \beta$ is fairly large, in which case the electroweak scale is given by
\begin{equation}
M_Z^2 = -2\left( m_{Hu}^2 + \left|\mu\right|^2 \right)+ \mathcal{O}\left(\frac{1}{\tan^2\beta}\right) \label{eq:mz}.
\end{equation}
In particular, we want to know the variation in the electroweak VEV as the UV parameters of the theory are varied in  a natural SUSY theory. 

Consider the dependence on the first two generation sfermion masses. If we take just the 2-loop expression for the beta function of the up type Higgs mass, arising from SU(2) and U(1) gauge interactions,
\begin{equation}
\frac{d m_{Hu}^2}{dt} \supset \frac{2}{\pi^2} \left( \sum \alpha_i^2\left(t\right) C_i\left(H_u\right)\right) \tilde{m}_{1,2}^2  \, , \label{eq:2loop}
\end{equation}
we would obtain a contribution containing a single logarithm,  giving a fairly small (but not completely negligible) tuning if $ \tilde{m}_{1,2}$ is of order a few TeV.\footnote{The other two loop contributions are all proportional to the yukawas of the first two generation fermions squared, and are completely negligible.} However, we know that the electroweak VEV has a strong dependence on the stop mass, which itself has a significant dependence on $ \tilde{m}_{1,2}$, hence is clearly associated to a tuning. The appropriate way to measure this is through a total derivative (where  $Z_{\tilde{m}_{1,2}^2} $ is defined as the fine tuning with respect to the UV value of $\tilde{m}_{1,2}^2$ )
\begin{align}
Z_{\tilde{m}_{1,2}^2} &= \frac{d\left(\log M_Z^2\right)}{d\left(\log \tilde{m}_{1,2}^2\right)} ,
\end{align}
which includes the effect of the sfermions feeding into the stops which then feed into the up type Higgs. This gives a three loop, logarithm squared contribution which can be significant, especially if the mediation scale is high. The shift in the stop mass as a result of a change in sfermion mass depends on the energy scale. Therefore it is necessary to integrate over all energy scales to obtain the weak scale fluctuation in the Higgs mass,
\begin{align}
\Delta m_{Hu}^2\left(M_W\right)|_{M_3} &= \int_{t_{\Lambda}}^{t_{M3}} \Delta \beta_{m_{Hu}^2}\left(t\right)|_{M3}\, dt ,\\
&=  \int_{t_{\Lambda}}^{t_{M3}} \frac{\partial  \beta_{m_{Hu}^2}\left(t\right)}{\partial m_{\tilde{t}}^2\left(t \right)} \Delta m_{\tilde{t}}^2\left(t \right)+ \frac{\partial  \beta_{m_{Hu}^2}\left(t\right)}{\partial \tilde{m}_{1,2}^2\left(t \right)} \Delta \tilde{m}_{1,2}^2\left(t \right) ,
\end{align}
where the first term is the contribution through the stop, and the second is the direct two loop contribution. The fine tuning is then given by (using the approximate relation between the weak scale up type Higgs mass and the Z mass \eqref{eq:zuphiggs})
\begin{align}
\frac{d\left(\log M_Z^2\right)}{d\left(\log \tilde{m}_{1,2}^2\right)} &=  \frac{\partial}{\partial \log\left(\tilde{m}_{1,2}^2\right)} \int_{t_{\Lambda}}^{t_{m1,2}} \frac{\partial\left( \frac{d}{dt}\left(\log\left(M_Z^2\right)\right)\right)}{\partial m_{\tilde{t}}^2\left(t \right)} m_{\tilde{t}}^2\left(t \right)+ \frac{\partial\left( \frac{d}{dt}\left(\log\left(M_Z^2\right)\right)\right)}{\partial  \tilde{m}_{1,2}^2\left(t \right)}  \tilde{m}_{1,2}^2\left(t \right)  dt\\
&= \frac{2 \tilde{m}_{1,2}^2}{M_Z^2}  \frac{\partial}{\partial \left(\tilde{m}_{1,2}^2\right)}  \int_{t_{\Lambda}}^{t_{m1,2}} \frac{\partial\left( \frac{d}{dt}m_{Hu}^2\right)}{\partial m_{\tilde{t}}^2\left(t \right)} m_{\tilde{t}}^2\left(t \right) + \frac{\partial\left( \frac{d}{dt}m_{Hu}^2\right)}{\partial \tilde{m}_{1,2}^2\left(t \right)} \tilde{m}_{1,2}^2\left(t \right)  dt  .
\end{align}
Using the expressions \eqref{eq:sft}, \eqref{eq:stoprunning} and \eqref{eq:2loop}, and including a factor to two to account for the fact that the coupling occurs through both the left and right handed stops, we obtain
\begin{align}
Z_{\tilde{m}_{1,2}^2} &=  \frac{\tilde{m}_{1,2}^2}{M_Z^2}  \frac{\partial}{\partial \left(\tilde{m}_{1,2}^2\right)}  \int_{t_{\Lambda}}^{t_{m12}}   \frac{3 m_t^2}{4 \pi^2 v^2 \cos\left(2 \beta \right)} \sum_i \frac{8C_i}{\pi b_i} \alpha_i \left( \frac{1}{1+\frac{b_i \alpha_i}{2\pi}\left(t_{\Lambda}-t\right)}-1\right) \tilde{m}_{1,2}^2 \\
& \qquad \qquad \qquad \qquad \qquad +  \frac{2}{\pi^2} \left( \sum \frac{\alpha_i^2 C_i\left(H_u\right)}{1+ \frac{b_i \alpha_i}{2\pi}\log\left(\frac{\Lambda_{UV}}{\tilde{m}_{1,2}}\right)} \right) \tilde{m}_{1,2}^2 \, dt \\[1em]
&= \frac{\tilde{m}_{1,2}^2}{M_Z^2} \frac{\partial}{\partial \left(\tilde{m}_{1,2}^2\right)} \sum_i \left[ A \frac{8 C_i}{\pi b_i} \alpha_i\left(\log\left(\frac{\Lambda}{\tilde{m}_{1,2}} \right)-\frac{2\pi}{\alpha_i b_i}\log\left(1+\frac{b_i \alpha_i}{2\pi}\log\left(\frac{\Lambda}{\tilde{m}_{1,2}} \right) \right)  \right) \right.\\
& \qquad \qquad \qquad \qquad \qquad \left. +   \frac{4 C_i}{\pi b_i} \alpha_i  \left( \frac{1}{1+\frac{b_i}{2\pi}\log\left(\frac{\Lambda_{UV}}{\tilde{m}_{1,2}\left(M_W\right)}\right)\alpha_i}-1\right)\right] \tilde{m}_{1,2}^2
, \label{eq:zm12}
\end{align}
where $A= \frac{3 m_t^2}{4 \pi^2 v^2 \cos\left(2 \beta \right)} $. As in the previous section, each term gives two contributions to the fine tuning. The largest  is from the direct variation of the initial sfermion masses, while the second contribution comes from varying the end point of the logarithm, and is somewhat smaller. 

Intuitively, it is clear what is occurring in the first term. There are two fine tunings occurring at different levels in the theory, the electroweak VEV is tuned by the mass of the stop, which is itself tuned by the first two generations. The first term in \eqref{eq:zm12} is effectively the result of multiplying these together, and weighing them by a factor less than 1 to take into account that the gluino only generates a change in the stop mass after some running has occurred. Numerical evaluation shows that the second term (the two loop direct contribution) typically gives a shift in the mass squared of $\sim 10 \div 50\%$ of the first term, and acts in the opposite direction reducing the overall fine tuning. This is because the direct contribution decreases the Higgs mass squared, while the indirect contribution decreases the stop mass squared resulting in a less negative Higgs mass squared.

\begin{figure}[t!]
\begin{center}
\includegraphics[width=110mm]{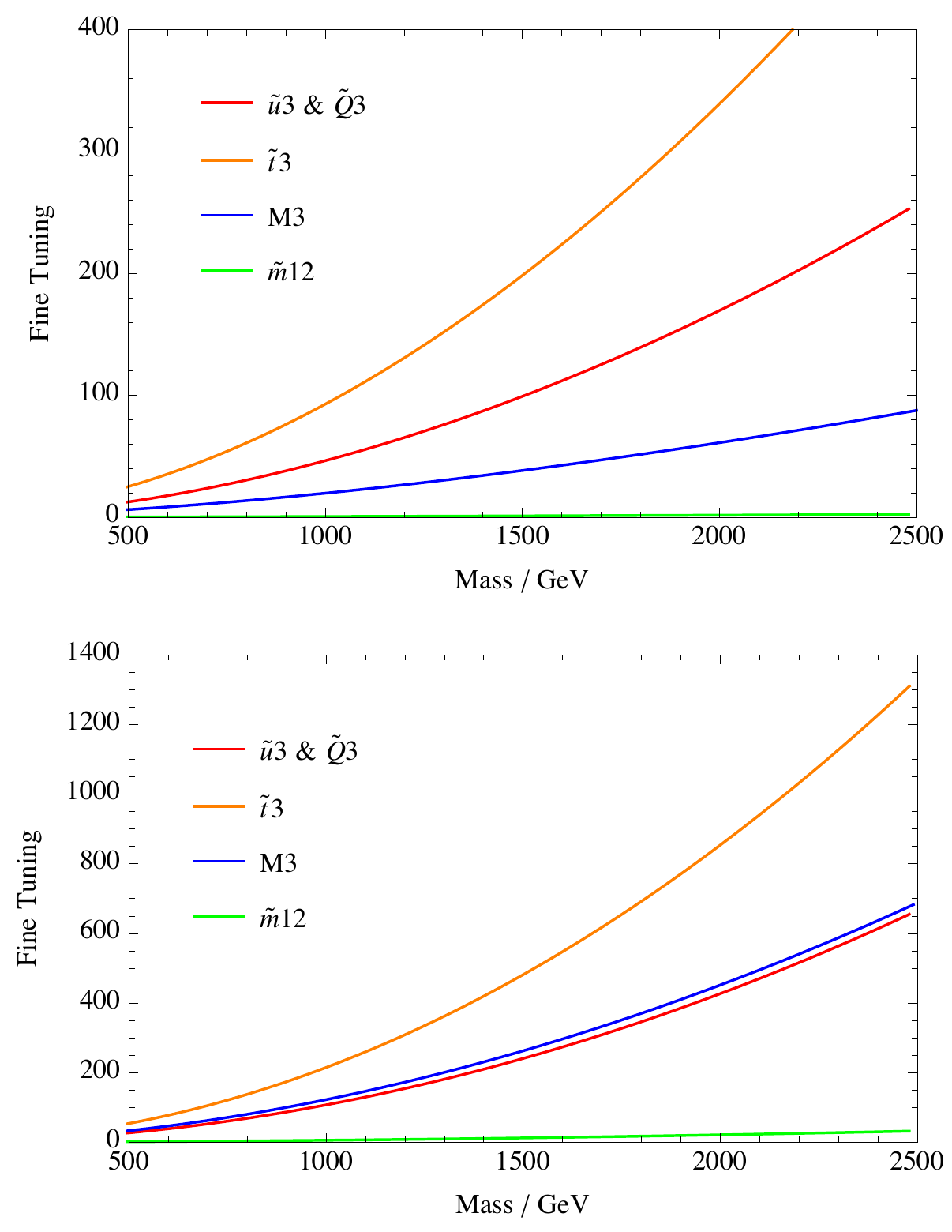}
\caption{The fine tuning in the electroweak sector as a function of the soft parameters, for low scale mediation with $\Lambda_{UV}=10^{6} \, \rm{GeV}$ (top) and high scale mediation $\Lambda_{UV}=10^{16} \, \rm{GeV}$ (bottom). The plots are a function of the \emph{weak scale} gluino mass since its running is fairly independent of the other parameters in the theory. The other masses are the values at the mediation scale, which may run to smaller or larger values when evolved to the weak scale.}
\label{fig:ewft}
\end{center}
\end{figure}

For the natural spectra we are interested in, the shift in the Higgs mass directly from the gluino is completely negligible compared to the logarithm squared contribution that occurs through the stop mass, hence we focus on the later.\footnote{The direct gluino contribution is two loop but only enhanced by a single logarithm compared to the two loop, two log enhanced contribution we study.} This gives
\begin{align}
Z_{M_3^2} &=  \frac{M_3^2}{M_Z^2} A \frac{\partial}{\partial \left(M_3^2\right)}  \int_{t_{\Lambda}}^{t_{M_3}}   \frac{4}{b_3} C_3 \left(\frac{1}{\left(1 + \frac{b_3 \alpha_3}{2\pi} \left(t_{\Lambda}-t\right)\right)^2}-1\right) M_3^2  dt  \\
&=  \frac{M_3^2}{M_Z^2} A \frac{\partial}{\partial \left(M_3^2\right)} \frac{4}{b_3} C_3 \frac{\frac{b_3 \alpha_3}{2\pi}\log^2\left(\frac{\Lambda}{M_3}\right)}{1+\frac{b_3 \alpha_3}{2\pi}\log\left(\frac{\Lambda}{M_3}\right)} M_3^2  . \label{eq:zm3}
\end{align}
Next, we turn to the tuning with respect to the initial stop mass. Since the renormalisation group equation governing the behaviour of a perturbation at the UV boundary of the stop mass may be solved exactly (at one loop order), as in \eqref{eq:stopt}, we can evaluate the shift in the low energy Higgs soft mass directly. This leads to
\begin{align}
\Delta m_{Hu}^2\left(M_Z\right) &= \frac{1}{2}\left( \left(\frac{m_{\tilde{Q3}}}{\Lambda_{UV}}\right)^{\left(3 y_t^2/4\pi^2 \right)}-1 \right) \Delta m_{\tilde{Q3}}^2\left(\Lambda_{UV}\right)\\
Z_{m_{\tilde{Q3}}^2} &=  \frac{m_{\tilde{Q3}}^2}{M_Z^2}  \frac{\partial}{\partial \left(m_{\tilde{Q3}}^2\right)} \left( \left(\frac{m_{\tilde{Q3}}}{\Lambda_{UV}}\right)^{\left(3 y_t^2/4\pi^2 \right)}-1 \right) \Delta m_{\tilde{Q3}}^2\left(\Lambda_{UV}\right) , \label{eq:zmu3}
\end{align}
for the left handed stop. The expression for the right handed stop is given by 
\begin{align}
Z_{m_{\tilde{u3}}^2}=   \frac{m_{\tilde{u3}}^2}{M_Z^2}  \frac{\partial}{\partial \left(m_{\tilde{u3}}^2\right)} \left( \left(\frac{m_{\tilde{u3}}}{\Lambda_{UV}}\right)^{\left(3 y_t^2/4\pi^2 \right)}-1 \right) \Delta m_{\tilde{u3}}^2\left(\Lambda_{UV}\right) . \label{eq:zmq3}
\end{align}
Alternatively, if we regard the UV masses of the left and right handed stops as one variable a similar computation easily gives
\begin{equation}
Z_{m_{\tilde{t3}}^2}= 2  \frac{m_{\tilde{t3}}^2}{M_Z^2}  \frac{\partial}{\partial \left(m_{\tilde{t3}}^2\right)}\left( \left(\frac{m_{\tilde{t3}}}{\Lambda_{UV}}\right)^{\left(3 y_t^2/4\pi^2 \right)}-1 \right) \Delta m_{\tilde{t3}}^2\left(\Lambda_{UV}\right)  . \label{eq:zt}
\end{equation}
If we assume the Higgs sector of the MSSM, solving the same set of renormalisation group equations, gives the tuning from a variation in the initial soft mass $m_{Hu}^2$ of
\begin{equation}
Z_{m_{Hu}^2}=   \frac{m_{Hu}^2}{M_Z^2}  \frac{\partial}{\partial \left(m_{Hu}^2\right)}\left( \left(\frac{m_{Hu}}{\Lambda_{UV}}\right)^{\left(3 y_t^2/4\pi^2 \right)}+1 \right) \Delta m_{Hu}^2\left(\Lambda_{UV}\right)  . \label{eq:ht}
\end{equation}
Assuming $\tan\beta$ is moderately sized, it is straightforward to check that the tuning with respect to $m_{Hd}^2$ is negligible compared to that from $m_{Hu}^2$. If the Higgs sector is more complicated, for example in the NMSSM, the exact expression here will be modified however it is still expected to still take the form $Z_{m_{Hu}^2} \lesssim 2 \frac{m_{Hu}^2}{M_Z^2} $, with the equality satisfied if $\Lambda_{UV}\sim m_{Hu}$ so there is very little running.

Finally,  we turn to the $\mu$ and $B\mu$ parameters. These do not feed strongly into other soft masses, and the tuning with respect to them is given by 
\begin{align}
Z_{\mu^2}=  2 \frac{\mu^2\left(\Lambda_{UV}\right)}{M_Z^2} \frac{\partial \mu^2\left(M_{Z}\right) }{\partial \mu^2\left(\Lambda_{UV}\right)} , \\
Z_{B\mu}=  2 \frac{B\mu\left(\Lambda_{UV}\right)}{M_Z^2} \frac{\partial B\mu\left(M_{Z}\right) }{\partial B\mu\left(\Lambda_{UV}\right)} ,
\end{align}
where the dependence of $M_Z$ on $B\mu$ arises from the terms which are higher order in $\frac{1}{\tan\beta}$. Assuming an MSSM Higgs sector and solving the renormalisation group equations numerically, we find that for $\mu = 400 \,\rm{GeV}$ and $B\mu = 200\,\rm{GeV}$ at the weak scale
\begin{align}
Z_{\mu^2} &\sim 40 ,\\
Z_{B\mu} & \sim 10  ,\\
\end{align}
for both high and low scale mediation. Since these values of $\mu$ and $B\mu$ are allowed by collider constraints, and it will turn out that the tunings are less than those from the stops, gluinos, and sfermions, the tunings from these parameters may be neglected from this point onwards. Once these parameters are fixed, the Higgs soft mass in the IR, and therefore after renormalisation flow at the UV boundary, is also fixed by \eqref{eq:mz}.\footnote{An alternative but equivalent approach would be to fix the UV boundary stop soft masses at a relatively small value in which case $\mu$ and $B\mu$ would be determined by the same relation.}

Having given expressions for the individual parameters at the renormalisation boundary scale, the overall fine tuning is taken to be $\Delta=\rm{max}\left(\{Z_i\}\right)$. Initially, we focus on the fine tuning introduced by the gluino mass, stop mass, and sfermion masses which are fairly independent of the exact details of the Higgs sector. In contrast, the fine tuning from the Higgs soft mass $m_{Hu}^2$ is dependent on both the $\mu$/$B\mu$ parameters, and whether the theory is the MSSM, the NMSSM, or some other extension (which is required in order to obtain the correct physical Higgs mass in some regions of parameter space). As a result, we study the fine tuning from $m_{Hu}^2$ in a typical MSSM Higgs sector seperately at the end of this Section. There it is seen that the tuning introduced is typically slightly smaller, but of the same order of magnitude, as that due to the other parameters. In addition, the regions with the lowest stop, gluino and sfermion fine tuning coincide with the regions where the Higgs has the lowest fine tuning. Therefore, the conclusions we draw about the overall tuning of the theory in the discussion that follows are valid, despite the omission of this important parameter.

Returning to the stop, gluino and sfermion soft masses, expanding the fine tuning expressions \eqref{eq:zm12}, \eqref{eq:zm3}  in the parameter $\frac{b_3 \alpha_3}{2\pi}\log\left(\frac{\Lambda_{UV}}{M_W}\right)$ and retaining only the leading dependence recovers the expressions used in previous papers such as \cite{Papucci:2011wy}.\footnote{Note this leads to a factor 2 difference in some expressions since we have included the finite energy range required for the  gluino to shift the stop mass.} However, since $\alpha_3$ is fairly large over all energy scales, and we are potentially interested in high scale models which can have large logarithms, we retain the full dependence in our numerical studies. In Fig.\ref{fig:ewft} we plot the fine tuning obtained as a result of the UV soft parameters for low and high scale mediation. We include both the cases where the stop masses are independent in the UV and when they are not. Not surprisingly, when they are both set by one parameter the fine tuning is rather worse since both feed into the up type Higgs mass simultaneously. 

\begin{figure}[t!]
\begin{center}
\includegraphics[width=125mm]{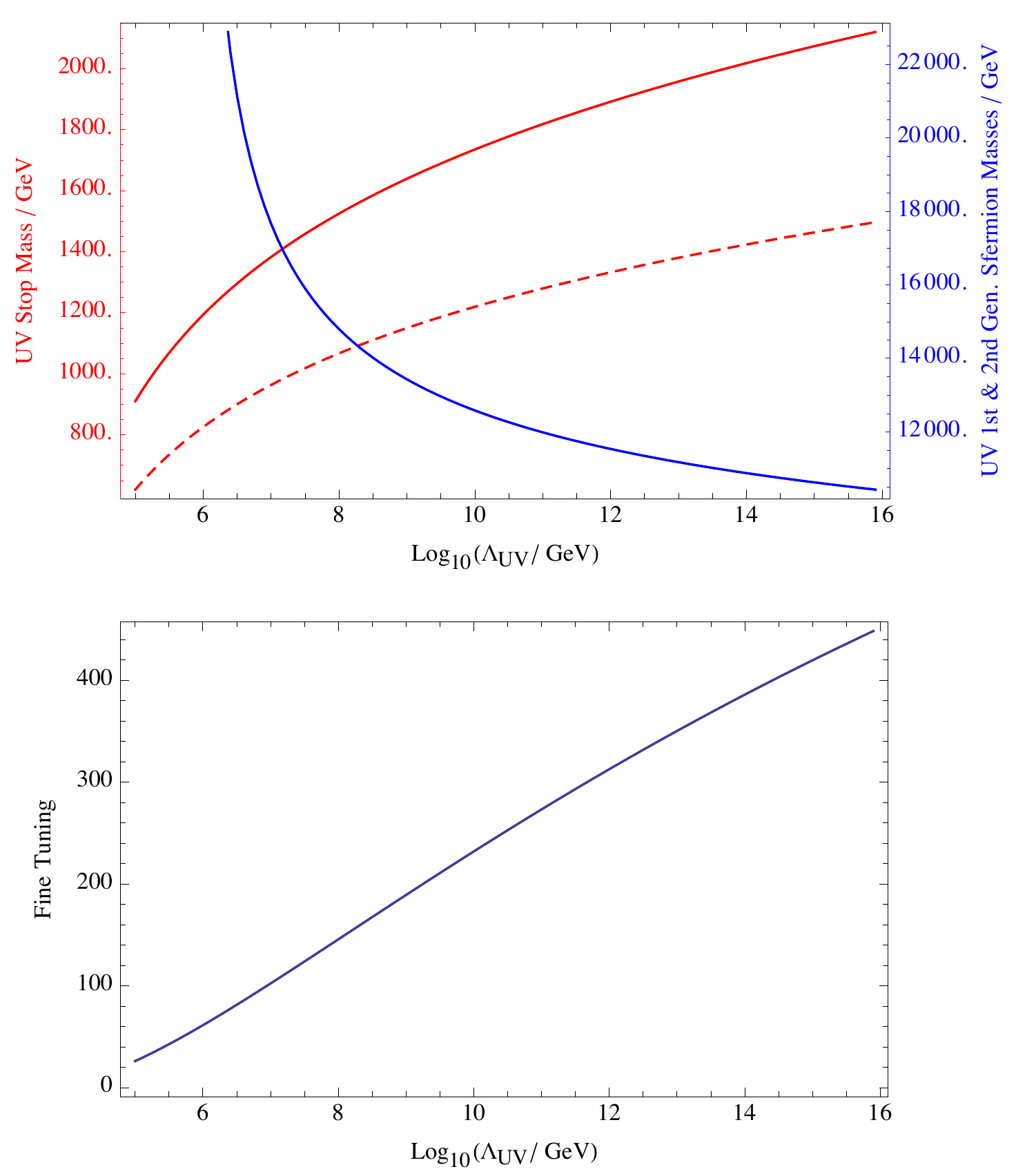}
\caption{{\bf Top:} The UV stop (red) and sfermion masses (blue) that lead to the same fine tuning of the electroweak scale as a gluino with weak scale mass of $2 \, \rm{TeV}$ as a function of the mediation scale. We show both the case where the left and right handed stops are independent parameters (solid lines) and when they are fixed equal (dashed). Lowering the stop or sfermion masses below these masses does not improve the fine tuning of the theory, and hence this graph limits the extent to which a natural spectrum can be obtained. {\bf Bottom:} The fine tuning corresponding to a $2 \, \rm{TeV}$ gluino as a function of mediation scale. By construction, this is the same as the fine tuning generated by stops at the masses in the top panel. If fine tuning better than 1\% is imposed then the mediation scale is limited to $\Lambda_{UV}<10^7\,\rm{GeV}$.}
\label{fig:massuv}
\end{center}
\end{figure}

The physics of these expressions is clear, for a given UV stop mass a larger gluino or sfermion mass is never actively favoured since they lead to greater fine tuning of the electroweak scale though their effect on the running of the stop.\footnote{It will be seen later that for a given \emph{weak} scale stop mass this does not necessarily hold as the UV stop mass is then a function of the gluino and sfermion masses.} However, provided $Z_{M^2_3}$ and $Z_{\tilde{m}_{1,2}^2}$ remain smaller than $Z_{m_{\tilde{t}}^2}$, increasing them does not actually make the fine tuning worse (at least with the measure of fine tuning adopted here). Hence, collider bounds can be somewhat alleviated without introducing fine tuning in the style of natural SUSY. It is interesting to ask what is the ratio of $m_{\tilde{t}}^2$, $M_{3}^2$, and $\tilde{m}_{1,2}^2$ which saturates a given fine tuning. In particular, suppose we fix the gluino mass to be $2 \, \rm{TeV}$ at the weak scale, we wish to know the maximum UV masses the stop and first two generation sfermions may have before they dominate the fine tuning. In Fig.\ref{fig:massuv} we plot the UV masses of the stops and first two generation sfermions for this scenario, both for the case of the left and right handed stops being independent, and when they are not. Hence, if the gluino is at $2 \, \rm{TeV}$, there is no fine tuning benefit to having UV stop masses below $1 \div 1.5 \, \rm{TeV}$ for GUT scale mediation, and  $0.5 \div 1 \, \rm{TeV}$ for very low scale mediation. From the bottom panel of Fig.\ref{fig:massuv} it is clear that a gluino of this mass forces the tuning of the electroweak scale to be at least $\sim 400$ if running begins at the GUT scale. In contrast, we see it is easily possible to separate the first two generation sfermions significantly from the gluino and stops without increasing the fine tuning of the theory, which is clearly beneficial for collider limits.


\begin{figure}[t!]
\begin{center}
\includegraphics[width=155mm]{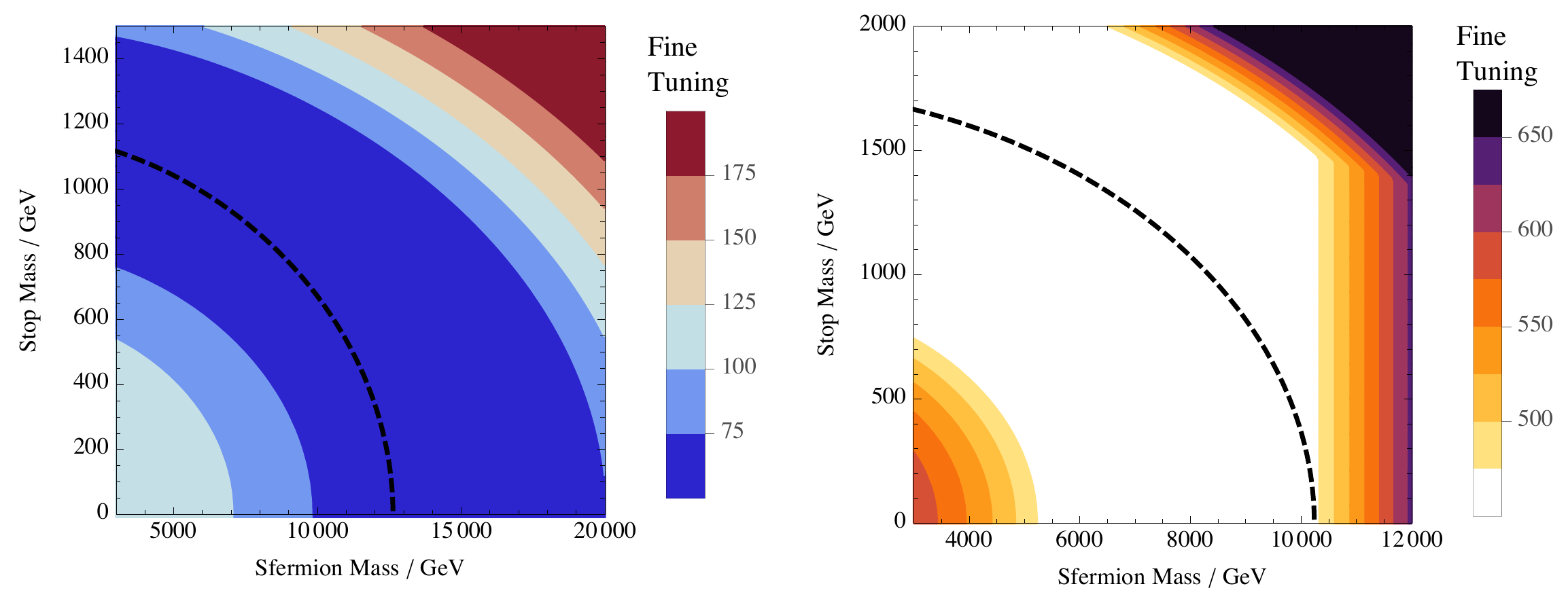}
\caption{The electroweak fine tuning as a function of the weak scale sfermion and stop masses (assuming the two stops are not independent) with a weak scale gluino mass of $2\,\rm{TeV}$ for {\bf Left:} $\Lambda_{UV}=10^{6}\,\rm{GeV}$, and {\bf Right:} $\Lambda_{UV}=10^{16}\,\rm{GeV}$. The regions below the dashed black line have a tachyonic stop mass at the UV boundary. Since sfermion masses $> 3\,\rm{TeV}$ are not constrained by collider limits, it is clear that for low scale mediation there is no improvement in fine tuning through decreasing the stops below $\sim 1.4 \,\rm{TeV}$. For high scale mediation, especially if we demand the stop is not tachyonic at the UV boundary, the majority of the region with the lowest fine tuning actually has a fairly heavy weak scale stop  $\sim 1.5 \,\rm{TeV}$.}
\label{fig:sfw}
\end{center}
\end{figure}

Of course, the relevant quantities for collider physics are the weak scale masses, and the running of the stops depends on the masses of the sfermions and the gluinos. As a result, the regions in Fig.\ref{fig:massuv} which minimise fine tuning with a given gluino mass can be somewhat shifted.  Therefore we plot the electroweak fine tuning as a function of the weak scale stop mass and first two generation sfermion masses with the weak scale gluino mass fixed at $2\,\rm{TeV}$,  for low and high scale mediation, in Fig.\ref{fig:sfw}.\footnote{The weak scale masses here are actually $\bar{\rm{MS}}$ masses and not pole masses. There is an additional finite correction to convert to the physical stop mass, but this is a small correction.} The conversion is carried out by numerically solving the renormalisation group equations between their UV boundary and the weak scale. It is assumed the two stops are not independent, however this does not qualitatively affect our conclusions. In these plots, due to the fixed gluino mass, the smallest possible electroweak fine tuning is $\sim 60$ and $\sim 450$ for low and high scale mediation respectively in agreement with  Fig.\ref{fig:massuv}. Hence, the large areas of parameter space with the lowest fine tuning in the centre of both plots have their fine tuning dominated by the gluino.

It is clear that for low scale mediation, there is no particular preference for the weak scale stop mass to be much lighter than $\lesssim 1.5\,\rm{TeV}$. For high scale mediation, the largest region of parameter space with low fine tuning actually has relatively large stop masses, $\sim 1.5\,\rm{TeV}$. In this case, heavy stop masses are even further favoured if we demand the stop is non-tachyonic at the boundary. This is a reasonable restriction since such boundary conditions can lead to deep colour breaking vacua in the early universe.\footnote{Although the existence of colour charge breaking vacua is not necessarily problematic if the color preserving vacua is metastable on timescales longer than the age of the universe \cite{Riotto:1995am,Riotto:1996xd,Kusenko:1996jn}, but there is still the problem of why our universe settled in the metastable vacuum during its early evolution.} As the sfermions tend to increase the stop mass during running up in energy,  the maximum weak scale stop mass that results in a tachyonic stop in the UV is decreased as the first two generation sfermions are made heavier.

In Fig.\ref{fig:sfw} contours of constant UV stop mass are approximately circle arcs concentric with the tachyonic contour. Starting from the line of tachyonic UV stops and moving outwards, the fine tuning starts to increase as new fine tuning contours are reached. This is the transition from the region dominated by the gluino tuning, to a region where the UV stop mass is the largest fine tuning. From Fig.\ref{fig:massuv}, this occurs when the UV stop mass is roughly $\sim 700\,\rm{GeV}$ and  $\sim 1000 \,\rm{GeV}$ for low and high scale mediation respectively.

On the far right side of the right panel, there is also a region where the sfermion controls the fine tuning, indicated by the vertical contours. Additionally, in the top right of both plots of Fig.\ref{fig:sfw}, there is a region where, if the sfermion mass was fixed, decreasing the stop mass would reduce the theory's fine tuning. However, since the LHC does not strongly constrain the first two generation sfermion masses in the ranges we are considering (in contrast to the gluino), we are not forced into this region of parameter space. Hence, the region with intermediate scale sfermion masses and relatively heavy stops, which has the lowest fine tuning possible for a fixed gluino mass, is favoured.

\begin{figure}[t!]
\begin{center}
\includegraphics[width=155mm]{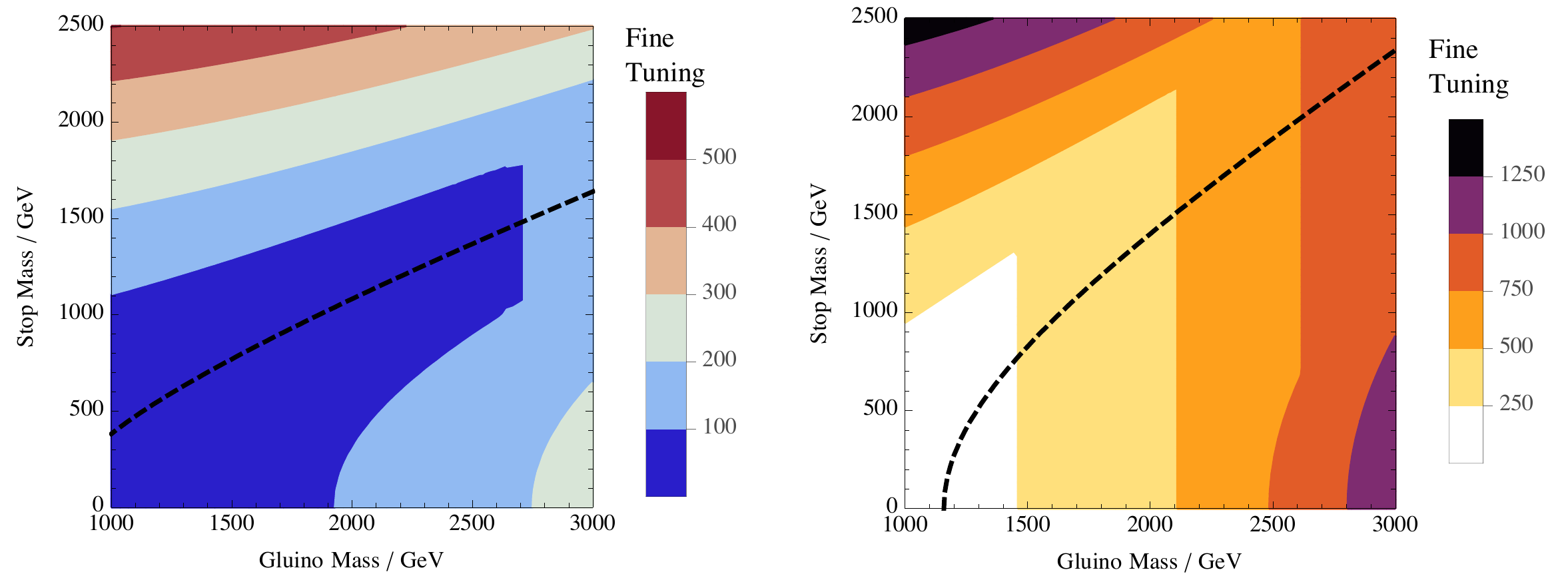}
\caption{The electroweak fine tuning as a function of the weak scale gluino and stop masses (assuming the two stops are not independent) with the sfermion mass fixed at 6 TeV for the cases {\bf Left:} $\Lambda_{UV}=10^{16}\,\rm{GeV}$ and  {\bf Right:} $\Lambda_{UV}=10^{6}\,\rm{GeV}$. The regions below the dashed black line have a tachyonic stop mass at the UV boundary.}
\label{fig:tmw}
\end{center}
\end{figure}

We also show the electroweak fine tuning as a function of the weak scale stop mass and gluino mass, with the first two generation sfermions fixed, in Fig.\ref{fig:tmw}. In these plots there is a lower bound on the fine tuning as a result of the fixed sfermion mass, which is reached in the  dark blue region in the left plot and the white region of the right plot. Considering a gluino mass of $\sim 2\,\rm{TeV}$ with high scale mediation, the contours are vertical for small stop masses. This is yet another sign that the gluino is dominating the fine tuning in this region, and the weak scale stop mass is unimportant provided it is $\lesssim 1.5\,\rm{TeV}$. In the tachyonic regions it can be seen that increasing the weak scale stop mass can actually improve the fine tuning. This occurs since increasing the stop mass leads to a less tachyonic UV boundary stop mass. As a result the ratio $\biggl|\frac{m_{\tilde{t}}^2\left(\Lambda_{UV} \right)}{m_{\tilde{t}}^2\left(M_W \right)}\biggr|$ is smaller, and the electroweak fine tuning is decreased if this is dominated by the stop mass. This can also be seen in the bottom left of both plots in  Fig.\ref{fig:sfw}.

For gluino masses of $\gtrsim 1.5\,\rm{TeV}$, the 6 TeV sfermions are far below their critical mass where significant tuning is introduced (for example by examining  Fig.\ref{fig:massuv}). As discussed, this is the most plausible scenario for natural spectra given LHC bounds, and the tuning is dominated by the stops or gluino. In these regions the upward pull of the gluino during running leads to fairly large stop masses, especially if we require the stop is not tachyonic at the UV boundary.


\begin{figure}[t!]
\begin{center}
\includegraphics[width=155mm]{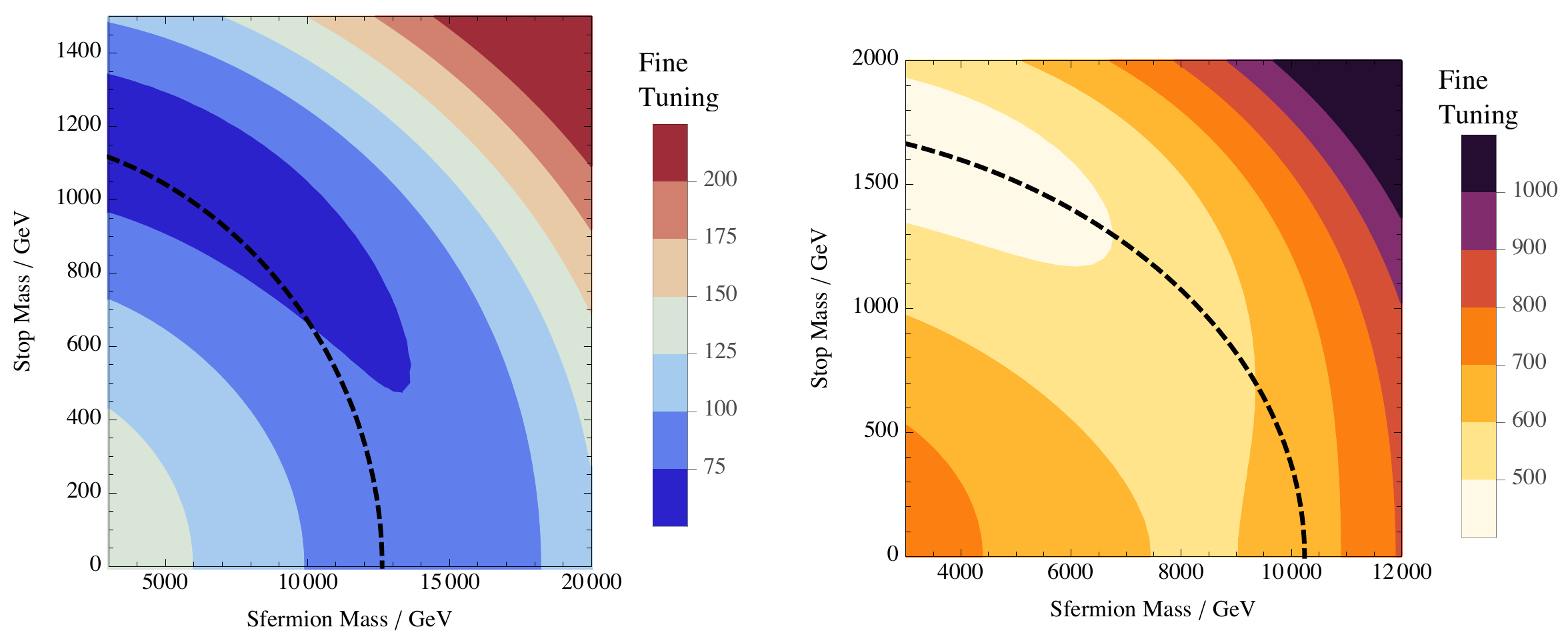}
\caption{The electroweak fine tuning, using the measure $\delta=\sqrt{\sum Z_i^2}$, as a function of the weak scale sfermion and stop masses (assuming the two stops are not independent) with a weak scale gluino mass of $2\,\rm{TeV}$ for {\bf Left:} $\Lambda_{UV}=10^{6}\,\rm{GeV}$, and {\bf Right:} $\Lambda_{UV}=10^{16}\,\rm{GeV}$. The regions below the dashed black line have a tachyonic stop mass at the UV boundary. As a result of the tuning introduced by the sfermions using this measure, lighter sfermions which correspond to heavier weak scale stops are favoured.}
\label{fig:sfwb}
\end{center}
\end{figure}

If instead an alternative definition of fine tuning,  $\delta= \sqrt{\sum_i Z_i^2}$, is used, broadly similar results are obtained. In this case, increasing the gluino or first two generation soft masses always increases the fine tuning, however until the critical points obtained above are reached this increase only makes the fine tuning worse by a modest amount. Above the critical soft masses, the fine tuning increases quickly as the soft masses are increased. In Fig.\ref{fig:sfwb} we plot the tuning as a function of the weak scale stop and first two generation sfermion masses, with the weak scale gluino fixed at $2\,\rm{TeV}$. The fine tuning pressure from the sfermion masses actually results in the regions with the smallest fine tuning having large stop masses.

Additionally, we briefly consider an alternative scenario discussed in the Introduction, where both the gluino and stop masses depend on the same F-term in the theory. In this case, $F^2$ is the fundamental parameter we should measure fine tuning with respect to. It is easy to see this scenario will be more fine tuned for a given gluino mass than our main case. Parametrically, the gluino mass is given by $M_3 \sim \frac{F^2}{M_{UV}^2}$ and the stop mass also by $m_{\tilde{t}}^2 \sim \frac{F^2}{M_{UV}^2}$ where $M_{UV}$ is a typical mass at the mediation scale. Hence, a $1\%$ increase in $F^2$ generates a $1\%$ increase in both the gluino and stop mass squared. As a result the fine tuning is worse than if the gluino and stop were independent variables. 

\begin{figure}[t!]
\begin{center}
\includegraphics[width=155mm]{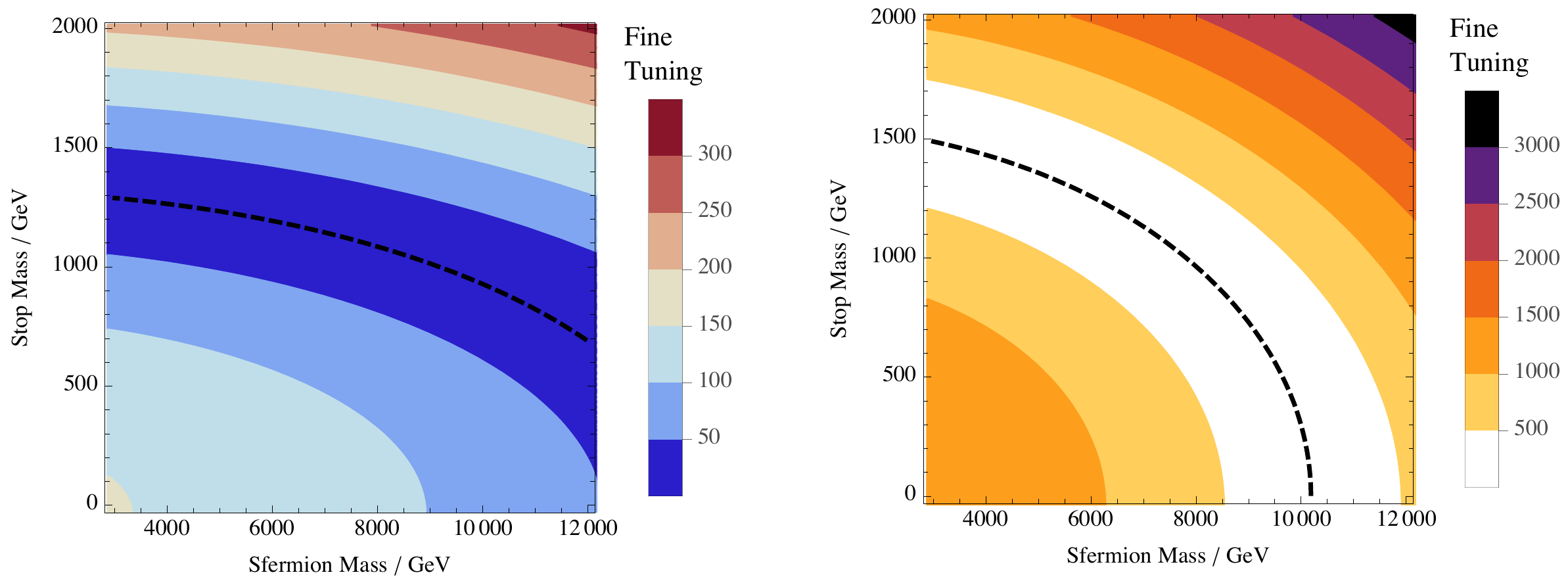}
\caption{The electroweak fine tuning from the initial up type Higgs mass, as a function of the weak scale sfermion and stop masses (assuming the two stops are not independent) with a weak scale gluino mass of $2\,\rm{TeV}$ for {\bf Left:} $\Lambda_{UV}=10^{6}\,\rm{GeV}$, and {\bf Right:} $\Lambda_{UV}=10^{16}\,\rm{GeV}$. The regions below the dashed black line have a tachyonic Higgs mass at the UV boundary.}
\label{fig:higgs}
\end{center}
\end{figure}

Finally, we return to the issue of the tuning as a result of the soft mass $m_{Hu}^2$. Taking $\mu= 400\,\rm{GeV}$, in Fig.\ref{fig:higgs} we plot this fine tuning as a function of the weak scale stop and sfermion masses, with the weak scale gluino fixed at $2\,\rm{TeV}$ in exact analogy to Fig.\ref{fig:sfw}. The fine tuning is calculated by numerically running the IR soft Higgs mass which gives the correct electroweak scale, to the UV boundary and evaluating \eqref{eq:ht}.\footnote{We assume vanishing A-terms although these may be important for generating the correct physical Higgs mass in some theories, and if large modify the running slightly.} Clearly, the tuning from the Higgs soft mass is not especially small. This is to be expected since the Higgs soft mass, of course, appears at tree level in the electroweak VEV. However, in the regions of lowest fine tuning,  the tuning from the Higgs mass is typically slightly smaller than that from the other parameters. The plot also shows that in the regions of lowest fine tuning the UV Higgs mass is not far near zero, and there are large parts of parameter space with small fine tuning where the Higgs soft mass squared (at the UV boundary of the renormalisation group flow) is positive. For low scale mediation, the part of parameter space with tuning less than $50$ has $\left| m_{Hu}^2\right| \lesssim \left(500\,\rm{GeV}\right)^2$ at the UV boundary of the RG flow, and for high scale mediation the region with tuning less than $500$ has  $\left| m_{Hu}^2\right| \lesssim \left(1000\,\rm{GeV}\right)^2$. The regions with low fine tuning actually coincide quite closely with the regions where the other parameters have low fine tuning. Therefore, our previous estimates of the fine tuning and favoured regions can be valid even when the details of a Higgs sector are included.

\section{Dirac Gauginos for Natural SUSY}\label{sec:dirac}
In this section we consider an interesting extension of the MSSM, Dirac gluinos. As is well known these provide an effective way of shielding the stop from gaining large corrections compared to the usual Majorana case. More precisely, in such theories there is an $\mathcal{N}=2$ supersymmetry in the gauge sector which means there are no infinite log enhanced corrections to the stop mass from the gluino. The only term is a finite piece generated below the scale where the heaviest part of the effective $\mathcal{N}=2$ multiplet is integrated out, which is typically the sgluon (the new scalar octet partner of the gluon), and above the mass of the gluino. As noted by many authors, Dirac gluinos provide a compelling mechanism for maintaining a supersymmetric spectrum without significant fine tuning \cite{Fox:2002bu,Kikuchi:2008ws,Carpenter:2010as,Davies:2011mp,Benakli:2011vb,Kribs:2012gx,Benakli:2012cy}. Hence, it is interesting to quantify the fine tuning obtained in such models.

We focus on a simple model, following \cite{Fox:2002bu,Kribs:2012gx}. There is an additional U(1) gauge group which obtains  a D-term expectation value, and has field strength $W'$. This couples to the visible sector $\mathcal{N}=2$ gauge multiplet, which can be written in $\mathcal{N}=1$ notation as a vector multiplet with field strength $W$, and a chiral multiplet $A$ in the adjoint of the gauge group, only through a term
\begin{equation}
\int d^2\theta \frac{\sqrt{2} W'_{\alpha}}{M_{UV}} W^{\alpha}_j A_j . \label{eq:dirac}
\end{equation}
It can be shown that this operator also induces a mass for the real component of the sgluon, $\tilde{m}_i^2$, of size $\tilde{m}_3=2 M_3$, where $M_i$ is the Dirac gaugino mass. \footnote{As discussed in  \cite{Fox:2002bu}, there actually exists another, independent, supersoft term coupling $W'$ and $A$ which gives a mass to the sgluon. For simplicity we assume this operator is absent from the theory.} In this minimal model there is no direct coupling between the SUSY breaking sector and the sfermions. Instead these are generated only by radiative corrections from the gauge sector as discussed in detail in \cite{Fox:2002bu}. The induced stop soft mass is given by
 \begin{align}
 \Delta m_{\tilde{t}}^2 &= \sum_i \frac{C_i \alpha_i}{\pi}  M_i^2 \log\left(\frac{\tilde{m}_i^2}{M^2_i} \right) , \label{eq:diracstop} \\
 &=   \frac{C_3 \alpha_3 M_3^2}{\pi} \log\left(4 \right) ,
 \end{align}
 where we have included only the dominant gluino contribution. The up type Higgs receives a contribution to its mass from the stop which is only present in the running between the stop soft mass and the scale at which this mass is generated. Since the stop mass is generated only in the small energy range between the sgluon and gluino masses, it is a reasonable approximation to assume it is tuned on instantaneously at the gluino mass.\footnote{This assumption leads to an error in the size of the logarithm in \eqref{eq:diracstop} of $\sim \frac{1}{2}\log 2\sim 0.3$, where the factor of $\frac{1}{2}$ is due to the finite energy range taken for the stop mass to be generated from the gluino mass. Since the typical value of the logarithm is $\log\left(\frac{M_3}{m_{\tilde{t}}}\right)\sim 2.5 $ this is negligable at the accuracy to which we are working.} Then the mass shift in the up type Higgs is given by
\begin{equation}
\Delta\left( \delta m_{Hu}^2\right) = -\frac{3 \lambda_t^2}{8 \pi^2} m_{\tilde{t}}^2 \log\left(\frac{M_3^2}{m^2_{\tilde{t}}} \right) ,
\end{equation}
which is clearly very suppressed relative to the MSSM case. Since the sgluon is heavier than the gluino, the energy range where the gluino mass feeds into the stop mass is separated from that in which the stop mass feeds into the Higgs mass, hence there is no need to carry out an integration over energies as we were required to do previously. The overall dependence of the Higgs mass on the gluino mass is then given by
\begin{align}
m_{Hu}^2|_{\rm{gluino}} &= \frac{-3 \lambda_t^2}{8\pi^2} m_{\tilde{t}}^2 \log\left(\frac{M_3^2}{m_{\tilde{t}}^2 } \right) \\
&=  \frac{3 \lambda_t^2}{8\pi^2}  \frac{C_3 \alpha_3 M_3^2}{\pi} \log\left(4 \right) \log\left(\frac{C_3 \alpha_3 }{\pi} \log\left(4 \right)  \right) .
\end{align}
Hence, the fine tuning is 
\begin{align}
\tilde{Z}_{M_3^2} &= \frac{M_3^2}{m_Z^2}  \frac{3 \lambda_t^2}{2\pi^2}  \frac{C_3 \alpha_3}{\pi} \log\left(4 \right) \log\left(\frac{C_3 \alpha_3 }{\pi} \log\left(4 \right)  \right) , \\
&\simeq  0.0282 \frac{M_3^2}{m_Z^2}  .
\end{align}
In this expression the two stop masses are not treated as independent variables since they are both generated through the gaugino masses, and cannot be adjusted independently. Therefore we regard the weak scale gluino mass as the only independent variable (or equivalently the stop mass). While, as previously discussed, using the weak scale value is an approximation, it is sufficient since there is very little running in such a theory. Further, since the running all occurs over a very small range of energies there is no need to account for the running of gauge couplings. Importantly, these expressions are independent of the mediation scale. This is due to the $\mathcal{N}=2$ structure cutting off the running at a lower scale, and this is what allows for the improvement in fine tuning.

The indirect fine tuning of the Higgs by the gluino through the stop mass, which was found to be the dominant contribution in the Majorana case, still appears as a logarithm squared, however now goes as
\begin{equation}
\sim  \log\left(\frac{M_3}{m_{\tilde{t}}} \right)  \log\left(\frac{\tilde{m}_i}{M_i} \right) ,
\end{equation}
which of course is much suppressed. In effect, the scale where a full $\mathcal{N}=2$ spectrum appears is acting as a UV boundary. This is a desirable alternative to a conventional model with a very low cutoff since it is still compatible with a string theory completion \cite{Davies:2012vu}, and avoids problematic higher dimension operators from a SUSY breaking and mediation sector which is not far separated in energy scale from the weak scale. Dirac models may also appear naturally out of models with spontaneous supersymmetry breaking \cite{Abel:2011dc}.

Since we are dealing with logarithms of $\mathcal{O}\left(1\right)$, such terms now no longer necessarily dominate over other finite, non-log enhanced corrections. To obtain an accurate measure of fine tuning these should be included. In particular, these are the reason that it is not possible to make the fine tuning arbitrarily small for heavy superpartners by taking $\tilde{m}_3=M_3$ and $M_3=m_{\tilde{t}}$. The threshold corrections from the gluino can be calculated from from \cite{Pierce:1996zz}. These make an $\mathcal{O}\left(1\right)$ difference to the stop masses generated. While it would be interesting to evaluate the full threshold corrections including those from the sgluon, in this section we consider only the log enhanced pieces as an approximation, leaving a full calculation to future work.


As the logarithms are now small, it is worthwhile to check the one loop contribution of the electroweakinos to the Higgs mass does not dominate the fine tuning. These give a contribution to the Higgs mass
\begin{equation}
\delta m_{Hu}^2= \delta m_{Hd}^2 = \frac{\alpha_2\left(M_2\right) C_2 M_2^2}{\pi} \log\left(\frac{\tilde{m}_2^2}{M_2^2}\right) ,
\end{equation}
which leads to a tuning of the electroweak VEV of approximately
\begin{align}
\Delta_{M2} &= \frac{M_2^2}{m_Z^2} \frac{2 \alpha_2\left(M_2\right) C_2 }{\pi}\left( \log\left(\frac{\tilde{m}_2^2}{M_2^2}\right)-1  \right)  ,\\
& = \frac{M_2^2}{m_Z^2} \frac{2 \alpha_2\left(M_2\right) C_2 }{\pi}\left( \log\left(4\right)-1  \right)  ,\\
&= 0.0062  \frac{M_2^2}{m_Z^2} .
\end{align}
Since the wino is typically significantly less massive than the gluino, this is only  a small contribution to the fine tuning.

\begin{figure}[t!]
\begin{center}
\includegraphics[width=120mm]{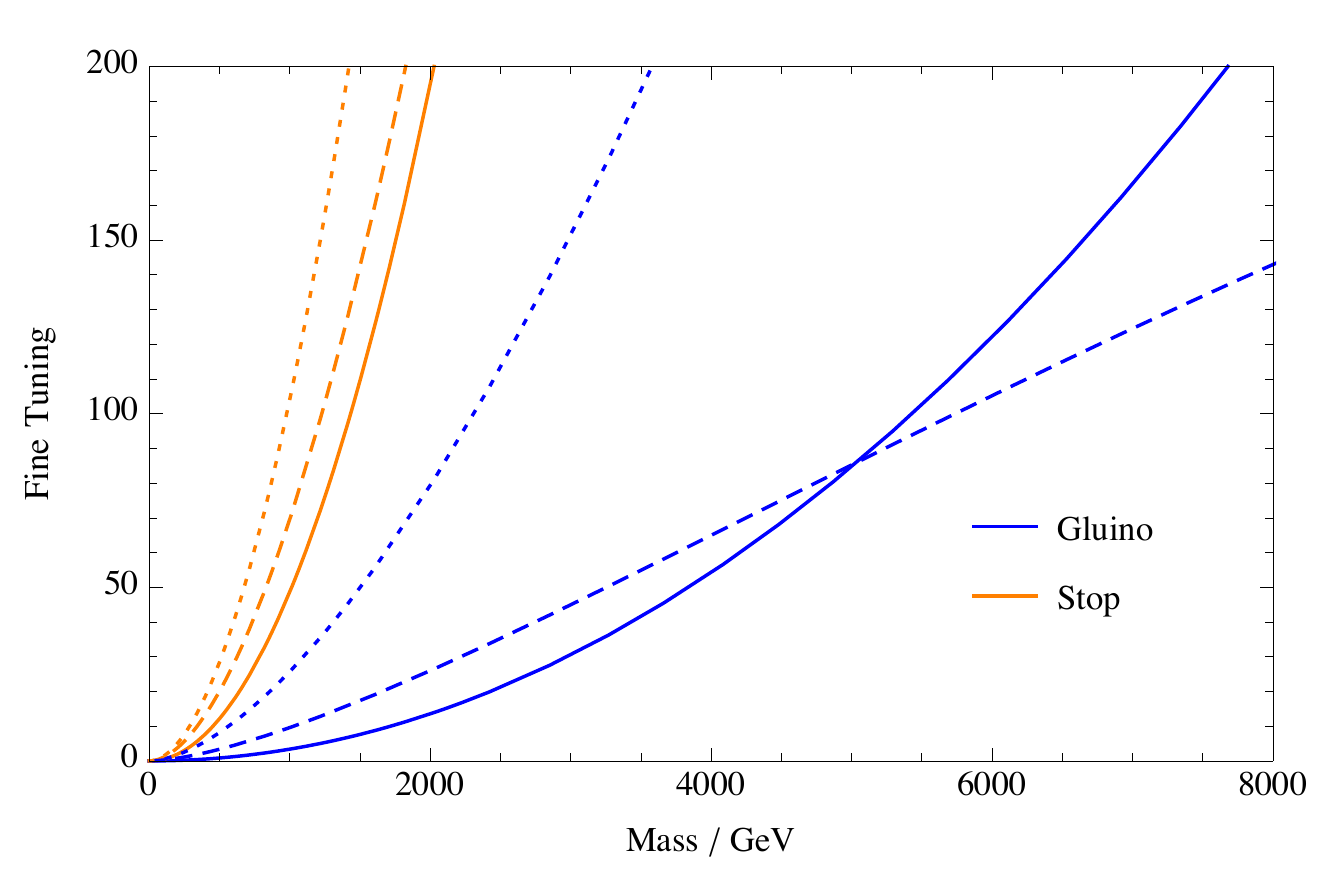}
\caption{The electroweak fine tuning of the minimal Dirac model as a function of gluino and stop masses (solid lines). Note that, in this model, the stop mass is a function of the gluino mass, hence these are not independent variables. For comparison we also plot the fine tuning for the MSSM, obtained in Section \ref{sec:ew}, for the cases $\Lambda_{UV}=10^5 \, \rm{GeV}$, dashed lines, and for $\Lambda_{UV}=10^6 \, \rm{GeV}$, dotted lines. It is seen that while the Dirac model gives comparable fine tuning to a very low scale MSSM model, it quickly leads to an improvement in fine tuning as the UV boundary is increased.}
\label{fig:dirac}
\end{center}
\end{figure}

In Fig.\ref{fig:dirac} we plot the fine tuning as a function of the gluino mass and also the stop mass which is fixed by the gluino mass (solid lines). However, as discussed, the accuracy of this is compromised by the small logarithms involved. The logarithm involved in the gluino generating the stop masses,  $\log\left(\frac{\tilde{m}_{3}^2}{M_3^2}\right)\sim 1.3$, is certainly not large enough to justify the neglect of the finite pieces. As a result, Fig.\ref{fig:dirac} should be regarded as giving a rough approximation for the fine tuning as a function of the gluino mass. In contrast, the logarithm involving the Higgs masses generated from the stops is given by $\log\left(\frac{M^2_3}{m^2_{\tilde{t}}}\right)\sim 5 $. Since this is somewhat larger,  the fine tuning as a function of stop mass in this figure is expected to be more accurate. 

By comparison with the expressions found in the previous section, we find the fine tuning as a function of stop mass is comparable to an MSSM model with a very low cutoff of $\Lambda_{UV}=10^5 \, \rm{GeV}$ (with both stops masses fixed by one parameter). However, as the cutoff $\Lambda_{UV}$ is raised, Dirac gauginos quickly lead to a benefit in reducing the fine tuning. Hence, for string models, a Dirac gluino provides a very strong option to retain as natural a spectrum as possible, as well as being well motivated theoretically. Of course, a disadvantage of such models is that the $\mathcal{N}=2$ scalar partners spoil traditional SUSY unification unless other new states are also present, requiring more model building.

\section{Summary}
We have carried out a careful study of the fine tuning in theories of natural supersymmetry, in particular concentrating on the tuning due to the UV masses of the gluino, first two generation sfermions, and the stops. In doing so, we have improved previous approximations which can introduce an $\mathcal{O}\left(1\right)$ correction to the results obtained. From these expressions, we have obtained limits on the extent to which it is beneficial to raise the gluino mass above the stop masses, and a lower bound on the fine tuning of theories for a given weak scale gluino mass. 

For models with high scale mediation,  if there is a Majorana gluino mass of $2 \,\rm{TeV}$ the fine tuning is at least $\simeq 400$, and only constrains the UV stop mass to be $\lesssim 1.5 \,\rm{TeV}$. After running to the weak scale, the stop mass can be up to $2 \,\rm{TeV}$ without affected the fine tuning, and in fact the largest regions of parameter space with the lowest fine tuning have fairly heavy IR stop masses of $\sim 1.5 \,\rm{TeV}$. Models with low scale mediation and a $2 \,\rm{TeV}$ Majorana gluino have a fine tuning of at least $\simeq 50$, and the UV stop mass is constrained to be $\lesssim 500 \,\rm{GeV}$. After running, the regions with the lowest fine tuning have IR stop masses up to $1400 \,\rm{GeV}$. In both high and low scale mediation models, the masses of the first two generation sfermions may be made very large, far out of reach of the LHC, without introducing additional fine tuning to the theory.

Finally, we have discussed an attractive alternative to the MSSM, Dirac gluinos. These allow for spectra with moderate fine and significant separation of the gluino and stops, comparable to that in theories with low scale mediation and a Majorana gluino, even if the scale of mediation is high. These are therefore a very attractive proposition for string-motivated models. 

\section*{Acknowledgements}
\noindent We are grateful to Asimina Arvanitaki, Masha Baryakhtar , Savas Dimopoulos, Saso Grozdanov, Ulrich Haisch, Xinlu Huang, Ken Van Tilburg, and James Unwin for useful discussions, and especially John March-Russell for discussions and comments on the manuscript. We are also grateful to the JHEP referee for a number of very useful comments, including on the distinction between an assumed UV boundary and a genuine UV cut-off and the importance of the direct two loop contribution from the sfermions to the Higgs.

\appendix

\section{Subleading Terms from Stop Back Reaction} \label{app:backreaction}
In this Appendix we calculate the back-reaction from the stop, when it is perturbed by a change in the gluino or sfermion masses. While this is a small effect, we include it in our numerical simulations. It occurs due to a term $\sim y_t m_{\tilde{Q3}}^2$ in the renormalisation group equation for the stops which tends to suppress any change in the stop mass. 

First, the effect on the gaugino fine tuning. If, at a scale t, a gaugino has led to a change in the left handed stop mass of $\Delta m_{\tilde{Q3}}^2$, this will feed back into the running as
\begin{equation}
\frac{d}{dt}\left(\Delta m_{\tilde{Q3}}^2 \right)= \frac{2 y_t^2}{16 \pi^2} \Delta m_{\tilde{Q3}}^2\left(t\right) .
\end{equation}
This can be integrated by using the expression for $\Delta m_{\tilde{Q3}}^2\left(t\right)$ in \eqref{eq:stoprunning}
\begin{align}
\Delta \left(\Delta m_{\tilde{Q3}}^2 \right) &= \frac{2 y_t^2}{16 \pi^2}\frac{2C_i}{b_i}\int_{t\Lambda}^{t} \left(M_i^2\left(t\right)-M_i^2\left(\Lambda\right)\right) dt \\
&= \frac{C_i y_t^2}{8\pi^2}\frac{\alpha_i\left(\Lambda\right)\log^2\left(\frac{\Lambda}{m_{\tilde{Q3}}}\right))}{1+\frac{b_i \alpha_i\left(\Lambda\right)}{2\pi}\log\left(\frac{\Lambda}{m_{\tilde{Q3}}}\right)} M_i^2 .
\end{align}
The leading expression for $\Delta \left(\Delta m_{\tilde{u3}}^2 \right)= 2\times \Delta \left(\Delta m_{\tilde{Q3}}^2 \right)$.

A similar procedure gives the back reaction from the first two generation sfermions as
\begin{equation}
\Delta \left(\Delta m_{\tilde{Q3}}^2 \right) = \sum_i \frac{y_t^2}{\pi^3 b_i} C_i  \alpha_i\left(\log\left(\frac{\Lambda}{m_{\tilde{Q3}}}\right) - \frac{2\pi}{b_i \alpha_i}\log\left(1+\frac{b_i\alpha_i}{2\pi}\log\left(\frac{\Lambda}{m_{\tilde{Q3}}}\right)\right)\right) \tilde{m}_{1,2}^2 ,
\end{equation}
and again $\Delta \left(\Delta m_{\tilde{u3}}^2 \right)= 2\times \Delta \left(\Delta m_{\tilde{Q3}}^2 \right)$. Since these corrections depend linearly on $M_i^2$ and $\tilde{m}_{1,2}^2$, their contribution to the fine tuning with respect to these variables is straightforward.

\bibliography{natural}

\providecommand{\href}[2]{#2}\begingroup\raggedright\begin{thebibliography}{10}

\bibitem{Dimopoulos:1995mi}
S.~Dimopoulos and G.~Giudice, {\it {Naturalness constraints in supersymmetric
  theories with nonuniversal soft terms}},  {\em Phys.Lett.} {\bf B357} (1995)
  573--578, [\href{http://xxx.lanl.gov/abs/hep-ph/9507282}{{\tt
  hep-ph/9507282}}].

\bibitem{Cohen:1996vb}
A.~G. Cohen, D.~Kaplan, and A.~Nelson, {\it {The More minimal supersymmetric
  standard model}},  {\em Phys.Lett.} {\bf B388} (1996) 588--598,
  [\href{http://xxx.lanl.gov/abs/hep-ph/9607394}{{\tt hep-ph/9607394}}].

\bibitem{Kribs:2013lua}
G.~D. Kribs, A.~Martin, and A.~Menon, {\it {Natural Supersymmetry and
  Implications for Higgs physics}},
  \href{http://xxx.lanl.gov/abs/1305.1313}{{\tt arXiv:1305.1313}}.

\bibitem{Krizka:2012ah}
K.~Krizka, A.~Kumar, and D.~E. Morrissey, {\it {Very Light Scalar Top Quarks at
  the LHC}},  \href{http://xxx.lanl.gov/abs/1212.4856}{{\tt arXiv:1212.4856}}.

\bibitem{Auzzi:2012dv}
R.~Auzzi, A.~Giveon, S.~B. Gudnason, and T.~Shacham, {\it {A Light Stop with
  Flavor in Natural SUSY}},  {\em JHEP} {\bf 1301} (2013) 169,
  [\href{http://xxx.lanl.gov/abs/1208.6263}{{\tt arXiv:1208.6263}}].

\bibitem{Espinosa:2012in}
J.~R. Espinosa, C.~Grojean, V.~Sanz, and M.~Trott, {\it {NSUSY fits}},  {\em
  JHEP} {\bf 1212} (2012) 077, [\href{http://xxx.lanl.gov/abs/1207.7355}{{\tt
  arXiv:1207.7355}}].

\bibitem{Han:2012fw}
Z.~Han, A.~Katz, D.~Krohn, and M.~Reece, {\it {(Light) Stop Signs}},  {\em
  JHEP} {\bf 1208} (2012) 083, [\href{http://xxx.lanl.gov/abs/1205.5808}{{\tt
  arXiv:1205.5808}}].

\bibitem{Lee:2012sy}
H.~M. Lee, V.~Sanz, and M.~Trott, {\it {Hitting sbottom in natural SUSY}},
  {\em JHEP} {\bf 1205} (2012) 139,
  [\href{http://xxx.lanl.gov/abs/1204.0802}{{\tt arXiv:1204.0802}}].

\bibitem{Bai:2012gs}
Y.~Bai, H.-C. Cheng, J.~Gallicchio, and J.~Gu, {\it {Stop the Top Background of
  the Stop Search}},  {\em JHEP} {\bf 1207} (2012) 110,
  [\href{http://xxx.lanl.gov/abs/1203.4813}{{\tt arXiv:1203.4813}}].

\bibitem{Allanach:2012vj}
B.~Allanach and B.~Gripaios, {\it {Hide and Seek With Natural Supersymmetry at
  the LHC}},  {\em JHEP} {\bf 1205} (2012) 062,
  [\href{http://xxx.lanl.gov/abs/1202.6616}{{\tt arXiv:1202.6616}}].

\bibitem{Larsen:2012rq}
G.~Larsen, Y.~Nomura, and H.~L. Roberts, {\it {Supersymmetry with Light
  Stops}},  {\em JHEP} {\bf 1206} (2012) 032,
  [\href{http://xxx.lanl.gov/abs/1202.6339}{{\tt arXiv:1202.6339}}].

\bibitem{Bi:2011ha}
X.-J. Bi, Q.-S. Yan, and P.-F. Yin, {\it {Probing Light Stop Pairs at the
  LHC}},  {\em Phys.Rev.} {\bf D85} (2012) 035005,
  [\href{http://xxx.lanl.gov/abs/1111.2250}{{\tt arXiv:1111.2250}}].

\bibitem{Brust:2011tb}
C.~Brust, A.~Katz, S.~Lawrence, and R.~Sundrum, {\it {SUSY, the Third
  Generation and the LHC}},  {\em JHEP} {\bf 1203} (2012) 103,
  [\href{http://xxx.lanl.gov/abs/1110.6670}{{\tt arXiv:1110.6670}}].

\bibitem{Arganda:2012qp}
E.~Arganda, J.~L. Diaz-Cruz, and A.~Szynkman, {\it {Decays of $H^0/A^0$ in
  supersymmetric scenarios with heavy sfermions}},  {\em Eur.Phys.J.} {\bf C73}
  (2013) 2384, [\href{http://xxx.lanl.gov/abs/1211.0163}{{\tt
  arXiv:1211.0163}}].

\bibitem{Arganda:2013ve}
E.~Arganda, J.~L. Diaz-Cruz, and A.~Szynkman, {\it {Slim SUSY}},  {\em
  Phys.Lett.} {\bf B722} (2013) 100,
  [\href{http://xxx.lanl.gov/abs/1301.0708}{{\tt arXiv:1301.0708}}].

\bibitem{Cao:2012rz}
J.~Cao, C.~Han, L.~Wu, J.~M. Yang, and Y.~Zhang, {\it {Probing Natural SUSY
  from Stop Pair Production at the LHC}},  {\em JHEP} {\bf 1211} (2012) 039,
  [\href{http://xxx.lanl.gov/abs/1206.3865}{{\tt arXiv:1206.3865}}].

\bibitem{ArkaniHamed:1997ab}
N.~Arkani-Hamed and H.~Murayama, {\it {Can the supersymmetric flavor problem
  decouple?}},  {\em Phys.Rev.} {\bf D56} (1997) 6733--6737,
  [\href{http://xxx.lanl.gov/abs/hep-ph/9703259}{{\tt hep-ph/9703259}}].

\bibitem{Hisano:2000wy}
J.~Hisano, K.~Kurosawa, and Y.~Nomura, {\it {Natural effective supersymmetry}},
   {\em Nucl.Phys.} {\bf B584} (2000) 3--45,
  [\href{http://xxx.lanl.gov/abs/hep-ph/0002286}{{\tt hep-ph/0002286}}].

\bibitem{Barbieri198863}
R.~Barbieri and G.~Giudice, {\it Upper bounds on supersymmetric particle
  masses},  {\em Nuclear Physics B} {\bf 306} (1988), no.~1 63 -- 76.

\bibitem{Kitano:2006gv}
R.~Kitano and Y.~Nomura, {\it {Supersymmetry, naturalness, and signatures at
  the LHC}},  {\em Phys.Rev.} {\bf D73} (2006) 095004,
  [\href{http://xxx.lanl.gov/abs/hep-ph/0602096}{{\tt hep-ph/0602096}}].

\bibitem{Kitano:2005wc}
R.~Kitano and Y.~Nomura, {\it {A Solution to the supersymmetric fine-tuning
  problem within the MSSM}},  {\em Phys.Lett.} {\bf B631} (2005) 58--67,
  [\href{http://xxx.lanl.gov/abs/hep-ph/0509039}{{\tt hep-ph/0509039}}].

\bibitem{Antusch:2012gv}
S.~Antusch, L.~Calibbi, V.~Maurer, M.~Monaco, and M.~Spinrath, {\it
  {Naturalness of the Non-Universal MSSM in the Light of the Recent Higgs
  Results}},  {\em JHEP} {\bf 01} (2013) 187,
  [\href{http://xxx.lanl.gov/abs/1207.7236}{{\tt arXiv:1207.7236}}].

\bibitem{Feng:2013pwa}
J.~L. Feng, {\it {Naturalness and the Status of Supersymmetry}},
  \href{http://xxx.lanl.gov/abs/1302.6587}{{\tt arXiv:1302.6587}}.

\bibitem{Arvanitaki:2012ps}
A.~Arvanitaki, N.~Craig, S.~Dimopoulos, and G.~Villadoro, {\it {Mini-Split}},
  {\em JHEP} {\bf 1302} (2013) 126,
  [\href{http://xxx.lanl.gov/abs/1210.0555}{{\tt arXiv:1210.0555}}].

\bibitem{Strumia:1999fr}
A.~Strumia, {\it {Naturalness of supersymmetric models}},
  \href{http://xxx.lanl.gov/abs/hep-ph/9904247}{{\tt hep-ph/9904247}}.

\bibitem{Kane:1998im}
G.~L. Kane and S.~King, {\it {Naturalness implications of LEP results}},  {\em
  Phys.Lett.} {\bf B451} (1999) 113--122,
  [\href{http://xxx.lanl.gov/abs/hep-ph/9810374}{{\tt hep-ph/9810374}}].

\bibitem{Kang:2012sy}
Z.~Kang, J.~Li, and T.~Li, {\it {On Naturalness of the MSSM and NMSSM}},  {\em
  JHEP} {\bf 1211} (2012) 024, [\href{http://xxx.lanl.gov/abs/1201.5305}{{\tt
  arXiv:1201.5305}}].

\bibitem{Cassel:2010px}
S.~Cassel, D.~Ghilencea, and G.~Ross, {\it {Testing SUSY at the LHC:
  Electroweak and Dark matter fine tuning at two-loop order}},  {\em
  Nucl.Phys.} {\bf B835} (2010) 110--134,
  [\href{http://xxx.lanl.gov/abs/1001.3884}{{\tt arXiv:1001.3884}}].

\bibitem{Cassel:2009cx}
S.~Cassel, D.~Ghilencea, and G.~Ross, {\it {Testing SUSY}},  {\em Phys.Lett.}
  {\bf B687} (2010) 214--218, [\href{http://xxx.lanl.gov/abs/0911.1134}{{\tt
  arXiv:0911.1134}}].

\bibitem{Ghilencea:2013fka}
D.~Ghilencea, {\it {A new approach to Naturalness in SUSY models}},
  \href{http://xxx.lanl.gov/abs/1304.1193}{{\tt arXiv:1304.1193}}.

\bibitem{Athron:2013ipa}
P.~Athron, M.~Binjonaid, and S.~F. King, {\it {Fine Tuning in the Constrained
  Exceptional Supersymmetric Standard Model}},
  \href{http://xxx.lanl.gov/abs/1302.5291}{{\tt arXiv:1302.5291}}.

\bibitem{Cabrera:2008tj}
M.~Cabrera, J.~Casas, and R.~Ruiz~de Austri, {\it {Bayesian approach and
  Naturalness in MSSM analyses for the LHC}},  {\em JHEP} {\bf 0903} (2009)
  075, [\href{http://xxx.lanl.gov/abs/0812.0536}{{\tt arXiv:0812.0536}}].

\bibitem{Casas:2003jx}
J.~Casas, J.~Espinosa, and I.~Hidalgo, {\it {The MSSM fine tuning problem: A
  Way out}},  {\em JHEP} {\bf 0401} (2004) 008,
  [\href{http://xxx.lanl.gov/abs/hep-ph/0310137}{{\tt hep-ph/0310137}}].

\bibitem{deCarlos:1993yy}
B.~de~Carlos and J.~Casas, {\it {One loop analysis of the electroweak breaking
  in supersymmetric models and the fine tuning problem}},  {\em Phys.Lett.}
  {\bf B309} (1993) 320--328,
  [\href{http://xxx.lanl.gov/abs/hep-ph/9303291}{{\tt hep-ph/9303291}}].

\bibitem{Badziak:2012rf}
M.~Badziak, E.~Dudas, M.~Olechowski, and S.~Pokorski, {\it {Inverted Sfermion
  Mass Hierarchy and the Higgs Boson Mass in the MSSM}},  {\em JHEP} {\bf 1207}
  (2012) 155, [\href{http://xxx.lanl.gov/abs/1205.1675}{{\tt
  arXiv:1205.1675}}].

\bibitem{Ross:2011xv}
G.~G. Ross and K.~Schmidt-Hoberg, {\it {The Fine-Tuning of the Generalised
  NMSSM}},  {\em Nucl.Phys.} {\bf B862} (2012) 710--719,
  [\href{http://xxx.lanl.gov/abs/1108.1284}{{\tt arXiv:1108.1284}}].

\bibitem{Baer:2012up}
H.~Baer, V.~Barger, P.~Huang, A.~Mustafayev, and X.~Tata, {\it {Radiative
  natural SUSY with a 125 GeV Higgs boson}},  {\em Phys.Rev.Lett.} {\bf 109}
  (2012) 161802, [\href{http://xxx.lanl.gov/abs/1207.3343}{{\tt
  arXiv:1207.3343}}].

\bibitem{Papucci:2011wy}
M.~Papucci, J.~T. Ruderman, and A.~Weiler, {\it {Natural SUSY Endures}},  {\em
  JHEP} {\bf 1209} (2012) 035, [\href{http://xxx.lanl.gov/abs/1110.6926}{{\tt
  arXiv:1110.6926}}].

\bibitem{Agashe:1998zz}
K.~Agashe and M.~Graesser, {\it {Supersymmetry breaking and the supersymmetric
  flavor problem: An Analysis of decoupling the first two generation scalars}},
   {\em Phys.Rev.} {\bf D59} (1999) 015007,
  [\href{http://xxx.lanl.gov/abs/hep-ph/9801446}{{\tt hep-ph/9801446}}].

\bibitem{Wymant:2012zp}
C.~Wymant, {\it {Optimising Stop Naturalness}},  {\em Phys.Rev.} {\bf D86}
  (2012) 115023, [\href{http://xxx.lanl.gov/abs/1208.1737}{{\tt
  arXiv:1208.1737}}].

\bibitem{Ghilencea:2012qk}
D.~Ghilencea and G.~Ross, {\it {The fine-tuning cost of the likelihood in SUSY
  models}},  {\em Nucl.Phys.} {\bf B868} (2013) 65--74,
  [\href{http://xxx.lanl.gov/abs/1208.0837}{{\tt arXiv:1208.0837}}].

\bibitem{Baer:2012mv}
H.~Baer, V.~Barger, P.~Huang, D.~Mickelson, A.~Mustafayev, {\em et.~al.}, {\it
  {Post-LHC7 fine-tuning in the mSUGRA/CMSSM model with a 125 GeV Higgs
  boson}},  \href{http://xxx.lanl.gov/abs/1210.3019}{{\tt arXiv:1210.3019}}.

\bibitem{Baer:2013bba}
H.~Baer, V.~Barger, and M.~Padeffke-Kirkland, {\it {Electroweak versus high
  scale finetuning in the 19-parameter SUGRA model}},
  \href{http://xxx.lanl.gov/abs/1304.6732}{{\tt arXiv:1304.6732}}.

\bibitem{Dine:2004dv}
M.~Dine, P.~Fox, E.~Gorbatov, Y.~Shadmi, Y.~Shirman, {\em et.~al.}, {\it
  {Visible effects of the hidden sector}},  {\em Phys.Rev.} {\bf D70} (2004)
  045023, [\href{http://xxx.lanl.gov/abs/hep-ph/0405159}{{\tt
  hep-ph/0405159}}].

\bibitem{Cohen:2006qc}
A.~G. Cohen, T.~S. Roy, and M.~Schmaltz, {\it {Hidden sector renormalization of
  MSSM scalar masses}},  {\em JHEP} {\bf 0702} (2007) 027,
  [\href{http://xxx.lanl.gov/abs/hep-ph/0612100}{{\tt hep-ph/0612100}}].

\bibitem{Contino:1998nw}
R.~Contino and I.~Scimemi, {\it {The Supersymmetric flavor problem for heavy
  first two generation scalars at next-to-leading order}},  {\em Eur.Phys.J.}
  {\bf C10} (1999) 347--356,
  [\href{http://xxx.lanl.gov/abs/hep-ph/9809437}{{\tt hep-ph/9809437}}].

\bibitem{Kribs:2007ac}
G.~D. Kribs, E.~Poppitz, and N.~Weiner, {\it {Flavor in supersymmetry with an
  extended R-symmetry}},  {\em Phys.Rev.} {\bf D78} (2008) 055010,
  [\href{http://xxx.lanl.gov/abs/0712.2039}{{\tt arXiv:0712.2039}}].

\bibitem{Abdullah:2012tq}
M.~Abdullah, I.~Galon, Y.~Shadmi, and Y.~Shirman, {\it {Flavored Gauge
  Mediation, A Heavy Higgs, and Supersymmetric Alignment}},  {\em JHEP} {\bf
  1306} (2013) 057, [\href{http://xxx.lanl.gov/abs/1209.4904}{{\tt
  arXiv:1209.4904}}].

\bibitem{Perez:2012mj}
M.~J. Perez, P.~Ramond, and J.~Zhang, {\it {Mixing supersymmetry and family
  symmetry breakings}},  {\em Phys.Rev.} {\bf D87} (2013), no.~3 035021,
  [\href{http://xxx.lanl.gov/abs/1209.6071}{{\tt arXiv:1209.6071}}].

\bibitem{Mahbubani:2012qq}
R.~Mahbubani, M.~Papucci, G.~Perez, J.~T. Ruderman, and A.~Weiler, {\it {Light
  non-degenerate squarks at the LHC}},
  \href{http://xxx.lanl.gov/abs/1212.3328}{{\tt arXiv:1212.3328}}.

\bibitem{Galon:2013jb}
I.~Galon, G.~Perez, and Y.~Shadmi, {\it {Non-Degenerate Squarks from Flavored
  Gauge Mediation}},  \href{http://xxx.lanl.gov/abs/1306.6631}{{\tt
  arXiv:1306.6631}}.

\bibitem{Dvali:1996rj}
G.~Dvali and A.~Pomarol, {\it {Anomalous U(1) as a mediator of supersymmetry
  breaking}},  {\em Phys.Rev.Lett.} {\bf 77} (1996) 3728--3731,
  [\href{http://xxx.lanl.gov/abs/hep-ph/9607383}{{\tt hep-ph/9607383}}].

\bibitem{Nelson:1997bt}
A.~E. Nelson and D.~Wright, {\it {Horizontal, anomalous U(1) symmetry for the
  more minimal supersymmetric standard model}},  {\em Phys.Rev.} {\bf D56}
  (1997) 1598--1604, [\href{http://xxx.lanl.gov/abs/hep-ph/9702359}{{\tt
  hep-ph/9702359}}].

\bibitem{Kaplan:1998jk}
D.~E. Kaplan, F.~Lepeintre, A.~Masiero, A.~E. Nelson, and A.~Riotto, {\it
  {Fermion masses and gauge mediated supersymmetry breaking from a single
  U(1)}},  {\em Phys.Rev.} {\bf D60} (1999) 055003,
  [\href{http://xxx.lanl.gov/abs/hep-ph/9806430}{{\tt hep-ph/9806430}}].

\bibitem{Kaplan:1999iq}
D.~E. Kaplan and G.~D. Kribs, {\it {Phenomenology of flavor mediated
  supersymmetry breaking}},  {\em Phys.Rev.} {\bf D61} (2000) 075011,
  [\href{http://xxx.lanl.gov/abs/hep-ph/9906341}{{\tt hep-ph/9906341}}].

\bibitem{Craig:2012di}
N.~Craig, M.~McCullough, and J.~Thaler, {\it {Flavor Mediation Delivers Natural
  SUSY}},  {\em JHEP} {\bf 1206} (2012) 046,
  [\href{http://xxx.lanl.gov/abs/1203.1622}{{\tt arXiv:1203.1622}}].

\bibitem{Hardy:2013uxa}
E.~Hardy and J.~March-Russell, {\it {Retrofitted Natural Supersymmetry from a
  U(1)}},  \href{http://xxx.lanl.gov/abs/1302.5423}{{\tt arXiv:1302.5423}}.

\bibitem{Brignole:1997dp}
A.~Brignole, L.~E. Ibanez, and C.~Munoz, {\it {Soft supersymmetry breaking
  terms from supergravity and superstring models}},
  \href{http://xxx.lanl.gov/abs/hep-ph/9707209}{{\tt hep-ph/9707209}}.

\bibitem{Meade:2008wd}
P.~Meade, N.~Seiberg, and D.~Shih, {\it {General Gauge Mediation}},  {\em
  Prog.Theor.Phys.Suppl.} {\bf 177} (2009) 143--158,
  [\href{http://xxx.lanl.gov/abs/0801.3278}{{\tt arXiv:0801.3278}}].

\bibitem{Feng:1999zg}
J.~L. Feng, K.~T. Matchev, and T.~Moroi, {\it {Focus points and naturalness in
  supersymmetry}},  {\em Phys.Rev.} {\bf D61} (2000) 075005,
  [\href{http://xxx.lanl.gov/abs/hep-ph/9909334}{{\tt hep-ph/9909334}}].

\bibitem{Horton:2009ed}
D.~Horton and G.~Ross, {\it {Naturalness and Focus Points with Non-Universal
  Gaugino Masses}},  {\em Nucl.Phys.} {\bf B830} (2010) 221--247,
  [\href{http://xxx.lanl.gov/abs/0908.0857}{{\tt arXiv:0908.0857}}].

\bibitem{Martin:1997ns}
S.~P. Martin, {\it {A Supersymmetry primer}},
  \href{http://xxx.lanl.gov/abs/hep-ph/9709356}{{\tt hep-ph/9709356}}.

\bibitem{Ellwanger:2009dp}
U.~Ellwanger, C.~Hugonie, and A.~M. Teixeira, {\it {The Next-to-Minimal
  Supersymmetric Standard Model}},  {\em Phys.Rept.} {\bf 496} (2010) 1--77,
  [\href{http://xxx.lanl.gov/abs/0910.1785}{{\tt arXiv:0910.1785}}].

\bibitem{Barbieri:2006bg}
R.~Barbieri, L.~J. Hall, Y.~Nomura, and V.~S. Rychkov, {\it {Supersymmetry
  without a Light Higgs Boson}},  {\em Phys.Rev.} {\bf D75} (2007) 035007,
  [\href{http://xxx.lanl.gov/abs/hep-ph/0607332}{{\tt hep-ph/0607332}}].

\bibitem{Hardy:2012ef}
E.~Hardy, J.~March-Russell, and J.~Unwin, {\it {Precision Unification in lambda
  SUSY with a 125 GeV Higgs}},  {\em JHEP} {\bf 1210} (2012) 072,
  [\href{http://xxx.lanl.gov/abs/1207.1435}{{\tt arXiv:1207.1435}}].

\bibitem{Allanach:2001kg}
B.~Allanach, {\it {SOFTSUSY: a program for calculating supersymmetric
  spectra}},  {\em Comput.Phys.Commun.} {\bf 143} (2002) 305--331,
  [\href{http://xxx.lanl.gov/abs/hep-ph/0104145}{{\tt hep-ph/0104145}}].

\bibitem{Riotto:1995am}
A.~Riotto and E.~Roulet, {\it {Vacuum decay along supersymmetric flat
  directions}},  {\em Phys.Lett.} {\bf B377} (1996) 60--66,
  [\href{http://xxx.lanl.gov/abs/hep-ph/9512401}{{\tt hep-ph/9512401}}].

\bibitem{Riotto:1996xd}
A.~Riotto, E.~Roulet, and I.~Vilja, {\it {Preheating and vacuum metastability
  in supersymmetry}},  {\em Phys.Lett.} {\bf B390} (1997) 73--79,
  [\href{http://xxx.lanl.gov/abs/hep-ph/9607403}{{\tt hep-ph/9607403}}].

\bibitem{Kusenko:1996jn}
A.~Kusenko, P.~Langacker, and G.~Segre, {\it {Phase transitions and vacuum
  tunneling into charge and color breaking minima in the MSSM}},  {\em
  Phys.Rev.} {\bf D54} (1996) 5824--5834,
  [\href{http://xxx.lanl.gov/abs/hep-ph/9602414}{{\tt hep-ph/9602414}}].

\bibitem{Fox:2002bu}
P.~J. Fox, A.~E. Nelson, and N.~Weiner, {\it {Dirac gaugino masses and
  supersoft supersymmetry breaking}},  {\em JHEP} {\bf 0208} (2002) 035,
  [\href{http://xxx.lanl.gov/abs/hep-ph/0206096}{{\tt hep-ph/0206096}}].

\bibitem{Kikuchi:2008ws}
T.~Kikuchi, {\it {A Solution to the little hierarchy problem in a partly N=2
  extension of the MSSM}},  \href{http://xxx.lanl.gov/abs/0812.2569}{{\tt
  arXiv:0812.2569}}.

\bibitem{Carpenter:2010as}
L.~M. Carpenter, {\it {Dirac Gauginos, Negative Supertraces and Gauge
  Mediation}},  {\em JHEP} {\bf 1209} (2012) 102,
  [\href{http://xxx.lanl.gov/abs/1007.0017}{{\tt arXiv:1007.0017}}].

\bibitem{Davies:2011mp}
R.~Davies, J.~March-Russell, and M.~McCullough, {\it {A Supersymmetric One
  Higgs Doublet Model}},  {\em JHEP} {\bf 1104} (2011) 108,
  [\href{http://xxx.lanl.gov/abs/1103.1647}{{\tt arXiv:1103.1647}}].

\bibitem{Benakli:2011vb}
K.~Benakli, {\it {Dirac Gauginos: A User Manual}},  {\em Fortsch.Phys.} {\bf
  59} (2011) 1079--1082, [\href{http://xxx.lanl.gov/abs/1106.1649}{{\tt
  arXiv:1106.1649}}].

\bibitem{Kribs:2012gx}
G.~D. Kribs and A.~Martin, {\it {Supersoft Supersymmetry is Super-Safe}},  {\em
  Phys.Rev.} {\bf D85} (2012) 115014,
  [\href{http://xxx.lanl.gov/abs/1203.4821}{{\tt arXiv:1203.4821}}].

\bibitem{Benakli:2012cy}
K.~Benakli, M.~D. Goodsell, and F.~Staub, {\it {Dirac Gauginos and the 125 GeV
  Higgs}},  \href{http://xxx.lanl.gov/abs/1211.0552}{{\tt arXiv:1211.0552}}.

\bibitem{Davies:2012vu}
R.~Davies, {\it {Dirac gauginos and unification in F-theory}},  {\em JHEP} {\bf
  1210} (2012) 010, [\href{http://xxx.lanl.gov/abs/1205.1942}{{\tt
  arXiv:1205.1942}}].

\bibitem{Abel:2011dc}
S.~Abel and M.~Goodsell, {\it {Easy Dirac Gauginos}},  {\em JHEP} {\bf 1106}
  (2011) 064, [\href{http://xxx.lanl.gov/abs/1102.0014}{{\tt
  arXiv:1102.0014}}].

\bibitem{Pierce:1996zz}
D.~M. Pierce, J.~A. Bagger, K.~T. Matchev, and R.-j. Zhang, {\it {Precision
  corrections in the minimal supersymmetric standard model}},  {\em Nucl.Phys.}
  {\bf B491} (1997) 3--67, [\href{http://xxx.lanl.gov/abs/hep-ph/9606211}{{\tt
  hep-ph/9606211}}].

\end{thebibliography}\endgroup
\bibliographystyle{JHEP}

\end{document}